\documentclass{IEEEoj}
\usepackage[nocompress]{cite}
\usepackage{amsmath,amssymb,amsfonts}
\usepackage{algorithmic}
\usepackage{graphicx,color}
\usepackage{textcomp}
\usepackage{comment}
\usepackage{colortbl}

\usepackage[colorlinks,urlcolor=blue,linkcolor=blue,citecolor=blue,breaklinks=true]{hyperref}

\usepackage{tabularx, booktabs, pifont, makecell, xparse}
\usepackage{multirow}
\usepackage{multicol}
\usepackage{xcolor}
\usepackage{hhline}
\usepackage{array}
\newcolumntype{C}[1]{>{\centering\arraybackslash}m{#1}}
\newcolumntype{Z}{>{\arraybackslash}X}
\definecolor{Color_High}{RGB}{255,229,153}
\definecolor{Color_Medium}{RGB}{189,215,238}
\definecolor{Color_Low}{RGB}{198,239,206}

\def\BibTeX{{\rm B\kern-.05em{\sc i\kern-.025em b}\kern-.08em
    T\kern-.1667em\lower.7ex\hbox{E}\kern-.125emX}}
\AtBeginDocument{\definecolor{ojcolor}{cmyk}{0.93,0.59,0.15,0.02}}
\def\OJlogo{\vspace{-4pt}\hskip-4pt\includegraphics[height=18pt]{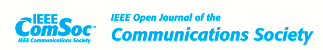}}
\begin{document}
\receiveddate{XX Month, XXXX}
\reviseddate{XX Month, XXXX}
\accepteddate{XX Month, XXXX}
\publisheddate{XX Month, XXXX}
\currentdate{XX Month, XXXX}
\doiinfo{OJCOMS.xxxx.xxxxxx}

\title{AI-Native Open RAN for Non-Terrestrial Networks: An Overview}

\author{JIKANG DENG\IEEEauthorrefmark{1} \IEEEmembership{(Student Member, IEEE)}, S.FIZZA HASAN\IEEEauthorrefmark{2}, HUI ZHOU\IEEEauthorrefmark{3} \IEEEmembership{(Member, IEEE)}, SAAD AL-AHMADI\IEEEauthorrefmark{2}, MOHAMED-SLIM ALOUINI\IEEEauthorrefmark{1}  \IEEEmembership{(Fellow, IEEE)}, AND DANIEL B. DA COSTA\IEEEauthorrefmark{2} \IEEEmembership{(Senior Member, IEEE)}}
\affil{CEMSE Division, King Abdullah University of Science and Technology (KAUST), Thuwal, Kingdom of Saudi
Arabia (KSA)}
\affil{Interdisciplinary Research Center for Communication Systems and Sensing (IRC-CSS), Department of Electrical Engineering, King Fahd University of Petroleum and Minerals (KFUPM), Dhahran,
KSA}
\affil{Centre for Future Transport and Cities, Coventry University, U.K.}
\corresp{CORRESPONDING AUTHOR: J. DENG (e-mail: jikang.deng@kaust.edu.sa).}
\markboth{AI-Native Open RAN for Non-Terrestrial Networks: An Overview}{Deng \textit{et al.}}

\begin{abstract}
Non-terrestrial network (NTN) is envisioned as a critical component of Sixth Generation (6G) networks by enabling ubiquitous services and enhancing network resilience. However, the inherent mobility and high-altitude operation of NTN pose significant challenges throughout the development and operations (DevOps) lifecycle. To address these challenges, integrating NTNs with the Open Radio Access Network (ORAN) is a promising approach, since ORAN can offer disaggregation, openness, virtualization, and embedded intelligence.
Despite extensive literature on ORAN and NTN, a holistic view of ORAN-based NTN frameworks is still lacking, particularly regarding how ORAN can effectively address the existing challenges of NTN. Furthermore, although artificial intelligence native (AI-Native) capabilities have the potential to enhance intelligence network control and optimization, their practical realization in NTNs has not yet been sufficiently investigated. 
Therefore, in this paper, we provide a comprehensive and structured overview of AI-Native ORAN for NTN. This paper commences with an in-depth review of the existing literature and subsequently introduces the necessary background about ORAN, NTN, and AI-Native for communication. After analyzing the DevOps challenges for NTN, we propose the orchestrated AI-Native ORAN-based NTN framework and discuss its key technological enablers. Finally, we present the representative use cases and outline the prospective future research directions of this study.
\end{abstract}


\maketitle

\section{INTRODUCTION}\label{sec:intro}
\subsection{MOTIVATION}
\IEEEPARstart{T}{he} emerging wireless networks, such as Beyond Fifth Generation (B5G) or Sixth Generation (6G) networks, are envisioned to merge the digital, physical, and human domains using intelligent infrastructures and platforms \cite{B5G-1}. Moreover, wireless networks are also expected to support a variety of applications and use cases that demand high data rates, low latency, and a very large number of users or devices with ubiquitous connectivity~\cite{B5G-2,B5G-4}. 

\begin{table*}[t]
    \centering
    \caption{List of abbreviations.}
    \label{tab:abbr}
    \renewcommand{\arraystretch}{1.1}
    \footnotesize
    \begin{tabular}{p{1.8cm} p{6cm} |p{1.8cm} p{6cm}}
        \toprule
        \textbf{Abbreviation} & \textbf{Definition} & \textbf{Abbreviation} & \textbf{Definition} \\
        \midrule\midrule
        5GC & Fifth-Generation Core & MIMO & Multiple-Input Multiple-Output \\
        6G & Sixth Generation & ML & Machine Learning \\
        AERPAW & Aerial Experimentation and Research Platform for Advanced Wireless & MNO & Mobile Network Operator \\
        AI & Artificial Intelligence & MOCN & Multiple Operator Core Network \\
        AIaaS & AI as a Service & Near-RT RIC & Near-Real-Time RAN Intelligent Controller \\
        B5G & Beyond Fifth Generation & NFV & Network Function Virtualization \\
        BS & Base Station & Non-RT RIC & Non-Real-Time RAN Intelligent Controller \\
        CapEx & Capital Expenditure & NTN & Non-Terrestrial Network \\
        CN & Core Network & NWDAF & Network Data Analytics Function \\
        COTS & Commercial Off-The-Shelf & O-CU & O-RAN Centralized Unit \\
        CSI & Channel State Information & O-DU & O-RAN Distributed Unit \\
        CTDE & Centralized Training and Distributed Execution & O-RU & O-RAN Radio Unit \\
        DevOps & Development and Operations & OpEx & Operating Expenditure \\
        DRL & Deep Reinforcement Learning & ORAN & Open Radio Access Network \\
        DT & Digital Twin & PDCP & Packet Data Convergence Protocol \\
        E2E & End-to-End & PHY & Physical Layer \\
        eMBB & Enhanced Mobile Broadband & RAN & Radio Access Network \\
        FH & Fronthaul & RF & Radio Frequency \\
        FL & Federated Learning & RIC & RAN Intelligent Controller \\
        FSO & Free-Space Optical & RLC & Radio Link Control \\
        GEO & Geostationary Earth Orbit & RRC & Radio Resource Control \\
        HAPs & High-Altitude Platforms & RT-RIC & Real-Time RAN Intelligent Controller \\
        ISAC & Integrated Sensing and Communication & SDAP & Service Data Adaptation Protocol \\
        LAE & Low-Altitude Economy & SMO & Service Management and Orchestration \\
        LEO & Low Earth Orbit & SWaP & Size, Weight, and Power \\
        LLMs & Large Language Models & TN & Terrestrial Network \\
        MAC & Medium Access Control & TPN & Transport Network \\
        MADRL & Multi-Agent Deep Reinforcement Learning & UAV & Uncrewed Aerial Vehicle \\
        MEO & Medium Earth Orbit & URLLC & Ultra-Reliable Low-Latency Communication \\
        \bottomrule
    \end{tabular}
\end{table*}

However, the deployment of dense 5G networks has revealed several limitations for vendors and mobile network operators (MNOs), including increased network complexity, high capital expenditure (CapEx), and operational expenditures (OpEx). In addition, energy efficiency has emerged as a critical issue in wireless networks. This is primarily due to the substantial energy consumption of the radio access network (RAN), which involves a dense deployment of base stations (BS) and large-scale multiple-input multiple-output (MIMO) systems at each site. More importantly, the coexistence of legacy and non-cloud-native applications, combined with users' requirements, has further highlighted the need for more intelligent, flexible, and cost-effective solutions in future 6G systems. Specifically, these challenges have motivated significant research into novel network architectures and communication schemes that aim to: (i) Meet the network performance requirements in terms of coverage, data rates, and latency; (ii) Support the envisioned applications such as integrated sensing and communication (ISAC), artificial intelligence (AI) agent-driven interactions, and cognitive cities; (iii) Enable the deployment of energy-efficient, scalable, and flexible networks; and (iv) Integrate intelligence and data to optimize the network performance and improve user experience with a variety of customized services.

The evolution of the RAN towards an open, disaggregated, and intelligent framework, i.e., Open RAN (ORAN), represents a key step in this direction \cite{openRAN-8}. Meanwhile, since B5G and 6G systems become increasingly integrated into important application scenarios, such as smart cities or healthcare, any service disruption can lead to serious socio-economic consequences, especially during emergencies or large-scale failures \cite{khaloopour2024resilience}. To extend coverage and enhance overall network resilience, the terrestrial network (TN) can be effectively complemented by a non-terrestrial network (NTN), which provides complementary and redundant communication links to achieve ubiquitous connectivity \cite{alves20256g,3gppntn}.
Nevertheless, the deployment and operation of NTNs, as well as their interoperability with TN, present several challenges. These include the size, weight, and power (SWaP) constraints, excessive propagation delays, and mobility management issues. These challenges have motivated increasing interest in the integration of ORAN and NTNs. This integration can leverage ORAN’s inherent features of disaggregation, openness, and AI-Native intelligence to enable more efficient network optimization and innovation \cite{dahrouj2023machine,rihan2023ran}. However, due to the aforementioned challenges, the adoption of ORAN principles in NTN remains significantly more complicated than in the traditional TN, thus requiring further studies on effective architectural and orchestration solutions.

\subsection{REVIEW OF EXISTING RELATED LITERATURE}
\label{state-of-the-art}
ORAN-based NTN has recently gained significant attention and research interests, with many research papers advancing the idea through investigations on different aspects. It should be noted that the scope of this subsection does not aim to cover the entirety of the existing works in ORAN for NTN, but rather to specifically review the research papers related to state-of-the-art development and deployment solutions of ORAN-based NTN. Therefore, a detailed content comparison between this survey paper and the most relevant and recent research papers is provided in Table \ref{table_comparison_related_work}.

\begin{table*}[!htb]
    \centering
    \footnotesize
    \renewcommand{\arraystretch}{1.3}
    \setlength{\tabcolsep}{4pt}
    \caption{Comparison of related works and this paper on ORAN-based NTN (\textbf{L: Low}, \textbf{M: Medium}, \textbf{H: High}).}
    \label{table_comparison_related_work}
    \begin{tabularx}{\textwidth}{|c|c|c|c|c|c|c|c|c|X|}
    \hline
    \rowcolor[HTML]{E0E0E0}
    \multicolumn{3}{|c|}{\textbf{Reference Info}} & 
    \multicolumn{6}{c|}{\textbf{ORAN-base NTN}} & 
    \textbf{Contribution} \\
    \hhline{|-|-|-|-|-|-|-|-|-|-|}
    \rowcolor[HTML]{E0E0E0}
    \textbf{Ref} & \textbf{Year}  & \textbf{Type} & \rotatebox{90}{\textbf{AI-Native}} & 
    \rotatebox{90}{\textbf{DevOps Challenges}} & 
    \rotatebox{90}{\textbf{Architecture}} & 
    \rotatebox{90}{\textbf{Key Enablers}} & 
    \rotatebox{90}{\textbf{Use cases}} & 
    \rotatebox{90}{\textbf{Future directions}} & 
    \\ 
    \hline
    \hline
    {\cite{baranda2025architectural}} & 2025 & \multirow{6}{*}{Survey} & \cellcolor{Color_Medium}M & \cellcolor{Color_Low}L & \cellcolor{Color_High}H & \cellcolor{Color_High}H & \cellcolor{Color_Low}L & \cellcolor{Color_Medium}M & 
    Analyzes functional splits and RIC placement for ORAN-enabled satellite-based NTNs, highlighting the implementation constraints and interoperability requirements.
    \\
    \hhline{|-|-|~|-|-|-|-|-|-|-|} 
    
    {\cite{openRAN-NTN-1}} & 2024 &  & \cellcolor{Color_High}H & \cellcolor{Color_Medium}M & \cellcolor{Color_High}H&
    \cellcolor{Color_Medium}M & \cellcolor{Color_Low}L & \cellcolor{Color_Medium}M & 
    Provides a detailed overview of the utilization of network slicing, AI/ML techniques, and ORAN to address the challenges of NTN systems in both academia and industry.
    \\
    \hhline{|-|-|~|-|-|-|-|-|-|-|} 

    {\cite{mushi2024open}} & 2024 &  & \cellcolor{Color_Low}L & \cellcolor{Color_Low}L & \cellcolor{Color_Medium}M & \cellcolor{Color_Medium}M & \cellcolor{Color_High}H & \cellcolor{Color_Medium}M & 
    Investigates the existing ORAN testbeds to support UAVs with controlled mobility, and evaluates AERPAW as a reference platform for future 6G NTN experimentation.
    \\
    \hhline{|-|-|-|-|-|-|-|-|-|-|}

    {\cite{mahboob2025transforming}} & 2025 & \multirow{19}{*}{Magazine} & \cellcolor{Color_Medium}M & \cellcolor{Color_Low}L & \cellcolor{Color_High}H & \cellcolor{Color_Medium}M & \cellcolor{Color_Low}L & \cellcolor{Color_Medium}M & 
    Proposes an ORAN-based NTN architectures with transparent or regenerative payloads configurations, and outlines the existing challenges in channel, computing and algorithms.
    \\
    \hhline{|-|-|~|-|-|-|-|-|-|-|}

    {\cite{baena2025space}} & 2025 &   & \cellcolor{Color_Medium}M & \cellcolor{Color_Medium}M & \cellcolor{Color_High}H & \cellcolor{Color_High}H & \cellcolor{Color_Medium}M  & \cellcolor{Color_Medium}M & 
    Proposes the Space-O-RAN framework, integrating satellite-based NTN and TN based on ORAN and AI-driven solutions, to address dynamic connectivity and scalability issues.
    \\
    \hhline{|-|-|~|-|-|-|-|-|-|-|}
    
    {\cite{firouzi20252}} & 2025 &  & \cellcolor{Color_Medium}M & \cellcolor{Color_Low}L & \cellcolor{Color_Medium}M & \cellcolor{Color_Medium}M & \cellcolor{Color_High}H & \cellcolor{Color_High}H & 
    Proposes the Orbital O-RAN (O$^2$-RAN) framework to support multi-layer, multi-segment NTN deployment, and enable intelligent and resilient operations across NTN infrastructures.
    \\
    \hhline{|-|-|~|-|-|-|-|-|-|-|}

    {\cite{shah2025urllc}} & 2025 &  & \cellcolor{Color_Medium}M & \cellcolor{Color_Low}L & \cellcolor{Color_Low}L &\cellcolor{Color_High}H & \cellcolor{Color_Low}L & \cellcolor{Color_Low}L& 
    Evaluates different functional split options in ORAN-enabled UAV-based NTNs for URLLC services in terms of computation complexity, latency, and reliability.
    \\ 
    \hhline{|-|-|~|-|-|-|-|-|-|-|}

    {\cite{zeydan2025semantic}} & 2025 &   & \cellcolor{Color_Medium}M & \cellcolor{Color_Low}L& \cellcolor{Color_High}H & \cellcolor{Color_Medium}M & \cellcolor{Color_Low}L & \cellcolor{Color_Medium}M & 
    Proposes the SEM-NTN framework, integrating semantic communication, ORAN, and NTNs, for real-time, bandwidth-constrained and AI-Native applications.
    \\
    \hhline{|-|-|~|-|-|-|-|-|-|-|}

    {\cite{moore2025next}} & 2025 &   & \cellcolor{Color_Medium}M & \cellcolor{Color_Low}L & \cellcolor{Color_Medium}M & \cellcolor{Color_Medium}M & \cellcolor{Color_Low}L & \cellcolor{Color_Medium}M & 
    Proposes an AERPAW-based ORAN experimentation platform prototype for wireless UAV research, and analyzes the key challenges and practical solutions for the platform.
    \\
    \hhline{|-|-|~|-|-|-|-|-|-|-|}
    
    {\cite{shinde2024ml}} & 2024 & & \cellcolor{Color_Medium}M & \cellcolor{Color_Low}L & \cellcolor{Color_Low}L & \cellcolor{Color_Medium}M & \cellcolor{Color_Low}L & \cellcolor{Color_Low}L & 
    Discusses the options and challenges for deploying ML solutions on ORAN-based TN and NTN scenarios to support centralized and distributed intelligent solutions.
    \\
    \hhline{|-|-|~|-|-|-|-|-|-|-|}

   {\cite{choi2024spectrum}} & 2024 &  & 
    \cellcolor{Color_Low}L & \cellcolor{Color_Low}L & \cellcolor{Color_Low}L & \cellcolor{Color_Medium}M & \cellcolor{Color_High}H & \cellcolor{Color_Low}L & 
    Discusses the utilization of ORAN architecture and spectrum marketplaces for facilitating spectrum-sharing strategies between NTNs and TN.    
    \\
    \hhline{|-|-|~|-|-|-|-|-|-|-|}

    {\cite{NTN-challenges-2}} & 2024&  & \cellcolor{Color_Low}L & \cellcolor{Color_Low}L & \cellcolor{Color_Medium}M & \cellcolor{Color_Low}L & \cellcolor{Color_Low}L & \cellcolor{Color_Medium}M & 
    Investigates the ORAN-based NTN architecture in enabling network optimization against various targets, and presents a neutral host 5G network testbed for evaluating the architecture.
    \\
    \hhline{|-|-|~|-|-|-|-|-|-|-|}

    {\cite{campana2023ran}} & 2023 &  & \cellcolor{Color_Low}L & \cellcolor{Color_Low}L & \cellcolor{Color_Medium}M & \cellcolor{Color_Low}L & \cellcolor{Color_Low}L & \cellcolor{Color_Medium}M & 
    Explores the implementation of ORAN-based NTN architecture and investigates the potential trends, such as radio resource management, to enhance NTN system efficiency.
    \\
    \hhline{|-|-|-|-|-|-|-|-|-|-|}

    {\cite{ORAN_white_paper}} & 2025 & White Paper  & \cellcolor{Color_Low}L & \cellcolor{Color_Medium}M & \cellcolor{Color_High}H & \cellcolor{Color_Medium}M & \cellcolor{Color_Medium}M & \cellcolor{Color_Low}L & 
    Presents a detailed overview of the integration of ORAN architecture and 3GPP NTN, emphasizing its importance, current status, challenges, and security considerations.
    \\
    \hhline{|-|-|-|-|-|-|-|-|-|-|}

    {\cite{pasumarthy2026designing}} & 2025 & Book Chapter & \cellcolor{Color_Low}L & \cellcolor{Color_Low}L & \cellcolor{Color_Medium}M & \cellcolor{Color_Medium}M & \cellcolor{Color_High}H & \cellcolor{Color_Low}L & 
    Summarizes the 3GPP-based NTN specifications, challenges, and various use cases, showing the benefits for realizing the NTN solutions based on ORAN architecture.
    \\
    \hhline{|-|-|-|-|-|-|-|-|-|-|}

    {\textbf{Ours}} & \textbf{2025} & Survey & \cellcolor{Color_High}\textbf{H} & \cellcolor{Color_High}\textbf{H} & 
    \cellcolor{Color_High}\textbf{H} & \cellcolor{Color_High}\textbf{H} & \cellcolor{Color_High}\textbf{H} & \cellcolor{Color_High}\textbf{H}& 
    \textbf{Provides a comprehensive overview of AI-Native
    ORAN-based NTN framework, highlighting the challenges, key enablers, use cases, and future research directions.}  \\
    \hline
    \end{tabularx}
\end{table*}

Some of these publications provide a detailed overview of the integration of ORAN and NTNs. For instance, the survey paper \cite{baranda2025architectural} investigated the architectural and functional split strategies and the RAN intelligent controller (RIC) placement options for ORAN-enabled satellite-based NTN systems, and emphasized the implementation constraints and interoperability considerations. Another in-depth survey paper \cite{openRAN-NTN-1} provided a detailed overview of the utilization of network slicing, AI or machine learning (ML) techniques, and ORAN to address the challenges of NTN systems in both academia and industry. Furthermore, the survey paper \cite{mushi2024open} discussed the design requirements of ORAN testbeds for controlled experimentation with UAV-based aerial clients, presented the representative experiments from aerial experimentation and research platform for advanced wireless (AERPAW) related to ORAN, and highlighted the AERPAW platform as a potential future ORAN-based UAV testbed.

Furthermore, several magazine papers focused on addressing the overall architecture related to ORAN-based NTNs \cite{mahboob2025transforming,baena2025space, firouzi20252}. Specifically, the authors in \cite{mahboob2025transforming} proposed an ORAN-empowered NTN architecture with transparent, regenerative-DU, and regenerative payload configurations, and the authors in \cite{baena2025space} proposed the Space-ORAN architecture framework by integrating satellite-based NTN and TN based on ORAN principles and AI-driven solutions. In \cite{firouzi20252}, a unified architectural framework named Orbital ORAN (O$^2$-RAN) is proposed to support multi-layer, multi-segment NTN deployment. In addition, other magazine papers have investigated specific aspects of ORAN-enabled NTNs, including ultra reliable low latency communications (URLLC) services \cite{shah2025urllc},  semantic communication \cite{zeydan2025semantic}, machine learning solution deployment \cite{shinde2024ml}, spectrum sharing \cite{choi2024spectrum}, and the implementation or testbed designs \cite{moore2025next, NTN-challenges-2,campana2023ran} based on AERPAW. Finally, the white paper \cite{ORAN_white_paper} analyzed the integration of ORAN architecture and 3GPP-based NTN systems in detail, and the book chapter \cite{pasumarthy2026designing} summarized the relevant characteristics for 3GPP NTN systems and highlighted the benefits of integrating NTN with ORAN architecture. Other surveys or magazine papers with limited relevance to the scope of ORAN for NTN can be found in \cite{NTORAN,muro20245g,sun2024advancing,soltani2025intelligent,openRAN-TN-1}.

Despite these extensive studies, as highlighted in Table \ref{table_comparison_related_work}, they had a limited focus and did not provide a holistic overview of ORAN-based NTN. For instance, while a few research papers had analyzed the architecture and key enablers in detail, they primarily focused on utilizing ORAN principles to enhance NTN in a similar way for TN, which lacks an in-depth analysis of the unique DevOps challenges for NTNs and results in simple technical adaptation. To obtain a broader and in-depth understanding of the necessity of ORAN-based NTN, it is essential to systematically investigate this topic, covering the existing NTN challenges, the benefits of AI-Native design, potential ORAN-based NTN architecture solutions, as well as representative use cases and future research directions.

\begin{figure*}[h!]
    \centering
    \includegraphics[width=\textwidth]{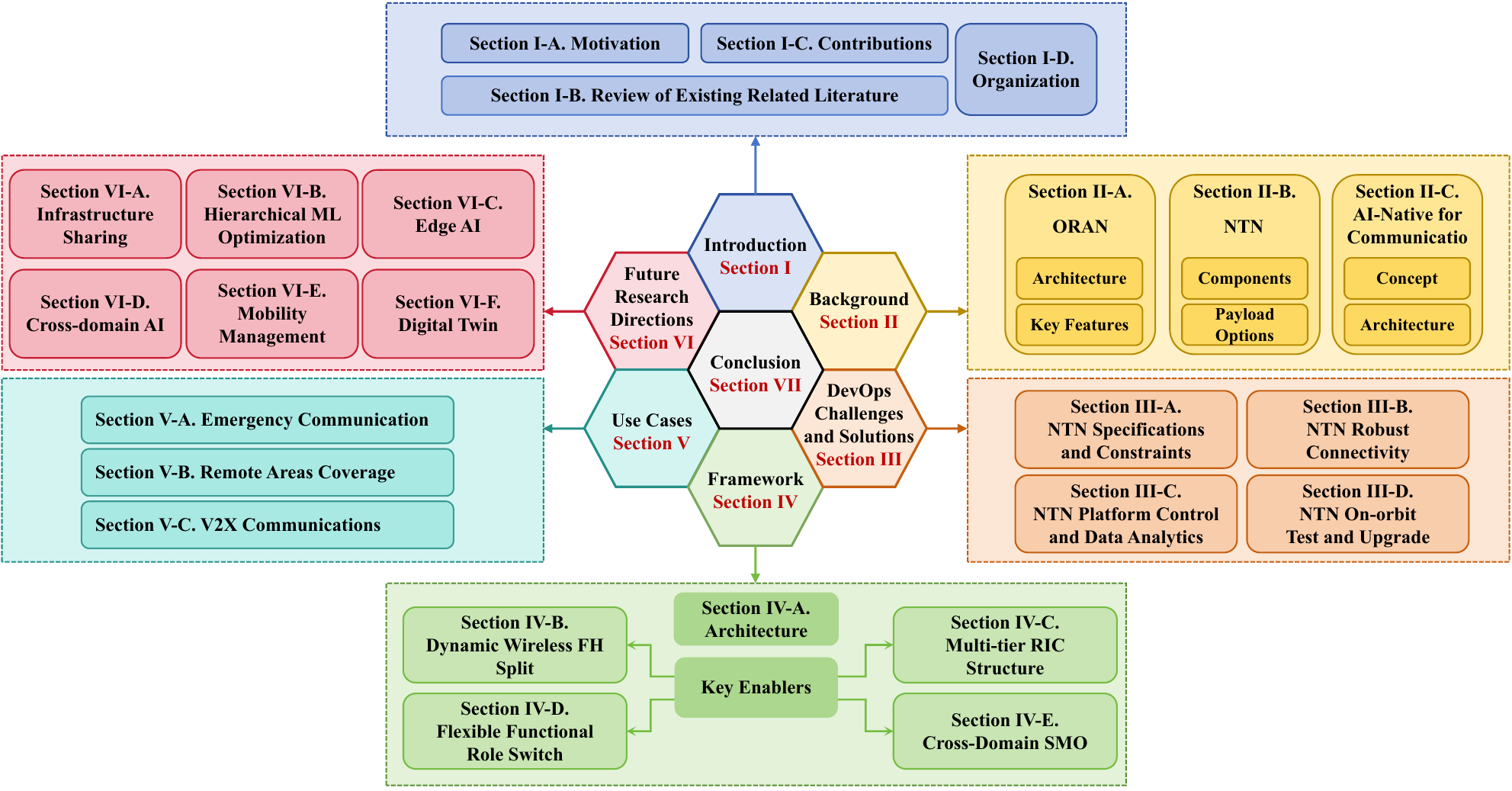}
    \caption{Detailed paper structure depicting the main sections of the paper and their organization.}
    \label{Fig_organization}
\end{figure*}

\subsection{CONTRIBUTIONS}
In this paper, we provide a comprehensive overview of the ORAN solutions for NTNs and analyze the potential of the AI-Native system in enhancing communication performance. Unlike related studies \cite{baranda2025architectural,openRAN-NTN-1,mushi2024open,mahboob2025transforming,baena2025space,firouzi20252,shah2025urllc,choi2024spectrum,zeydan2025semantic,shinde2024ml,moore2025next, NTN-challenges-2,campana2023ran,ORAN_white_paper,pasumarthy2026designing}, we analyze the unique DevOps challenges in the NTN lifecycle, propose an AI-Native ORAN-based NTN framework with key technological enablers, present various use cases, and discuss the future directions. The contributions of this paper are summarized as follows:
 \begin{itemize}
     \item We provide an up-to-date and structured review of the most recent and related studies on the ORAN-based NTNs, highlighting their current progress, contributions, and limitations. In addition, we provide a concise background on ORAN, NTN, and AI-Native for communication, and also present key survey and magazine papers for readers interested in exploring these topics further. 
     \item We identify and analyze the unique challenges across the DevOps lifecycle of NTNs, including specifications and constraints, robust connectivity, platform control and data analytics, and on-orbit test and upgrade. More importantly, we discuss the corresponding ORAN-based solutions by leveraging its inherent features to address these challenges. 
     \item We then propose a novel orchestrated AI-Native ORAN-based NTN framework to enable dynamic, scalable, and intelligent network operation for future 6G networks. We also highlight its key enabling technologies, including the dynamic wireless fronthaul (FH) split, multi-tier RIC structure, flexible functional role switch, and cross-domain service management and orchestration (SMO). 
     \item We present three representative use cases of the proposed framework, including emergency communication, remote areas coverage, and vehicle-to-everything (V2X) communications, which demonstrate the potential of the proposed ORAN-based NTN framework. 
     \item We discuss the future research directions on integrating the proposed ORAN-based NTN framework with other technologies, including infrastructure sharing, hierarchical ML-based optimization, edge AI, cross-domain AI, mobility management, and digital twin. These directions will further shape the evolution of AI-driven intelligent wireless networks. 
 \end{itemize}

\subsection{ORGANIZATION}
The remainder of this paper is organized as follows.
Section \ref{background} provides the detailed background for ORAN, NTN, and AI-Native for communication, and reviews the latest studies related to these topics.
Section \ref{NTN_devops_challenges} analyzes the unique challenges of NTN throughout the DevOps lifecycle and discusses corresponding ORAN-based solutions. 
Section \ref{ORAN_NTN_proposed} proposes a novel AI-Native ORAN-based NTN framework designed to achieve dynamic and intelligent network configuration, and introduces its key enabling technologies.
Section \ref{use cases} provides three representative use cases empowered by the proposed framework.
Section \ref{future directions} outlines future research directions for integrating the proposed framework with other emerging technologies.
Finally, Section~\ref{conclusion} concludes the paper. 
An overview of the paper organization is shown in Fig.~\ref{Fig_organization}, and the list of key abbreviations used is provided in Table~\ref{tab:abbr}.

\section{BACKGROUND} \label{background}
This section is intended to provide a brief background about ORAN, NTNs, and AI-Native for communication. We have listed some important and representative survey and/or magazine papers in each subsection for the interested readers to further explore these topics. 

\subsection{ORAN} 
In this sub-section, we aim to provide the readers with a clear understanding of the ORAN architecture and its key characteristics. 
A wide range of survey and tutorial articles have examined ORAN from various perspectives, providing valuable insights into its architecture, design principles, and implementation. Representative works \cite{openRAN-4, openRAN-TN-1, azariah2024survey, openRAN-7, openRAN-8} offered comprehensive overviews of ORAN development, interfaces, and open research challenges. In addition, other surveys and tutorials presented a broad overview of ORAN along with focused discussions on specific topics such as network slicing \cite{ORAN-survey-2025}, simulation and testbeds \cite{herrera2025tutorial}, and spectrum allocation \cite{martian2025towards}. The aforementioned surveys and tutorials can serve as valuable complementary resources for readers seeking further background on ORAN relevant to this study.

\subsubsection{ARCHITECTURE}
The RAN architectures have evolved from the conventional designs used in 2G to 4G networks toward virtualized RAN (vRAN) and cloud RAN (C-RAN), in which network functions and services are decoupled from dedicated hardware and implemented virtually \cite{openRAN-3}. This evolution has further led to ORAN, where multi-vendor components are disaggregated, virtualized, and connected through open and well-defined interfaces. 

The O-RAN Alliance was founded in 2018 to promote openness and intelligence in the RAN after the two organizations C-RAN Alliance and the xRAN forum joined forces. The main goal of the Alliance is to standardize an architecture and set of interfaces for the ORAN paradigm \cite{openRAN-2}. The alliance introduces the ORAN architecture and specifications to extend the RAN standards to include openness and intelligence. The alliance has 11 technical work groups (WGs) for the specification tasks in the ORAN architecture, 5 focus groups for the topics that are over-arching the technical WGs or are relevant for the whole organization, and a next Generation Research Group (nGRG) focusing on the research of open and intelligent RAN principles in 6G and future network standards \cite{openRAN-1}.

\begin{figure}[ht]
	\centering
    \includegraphics[width= 8.5cm]{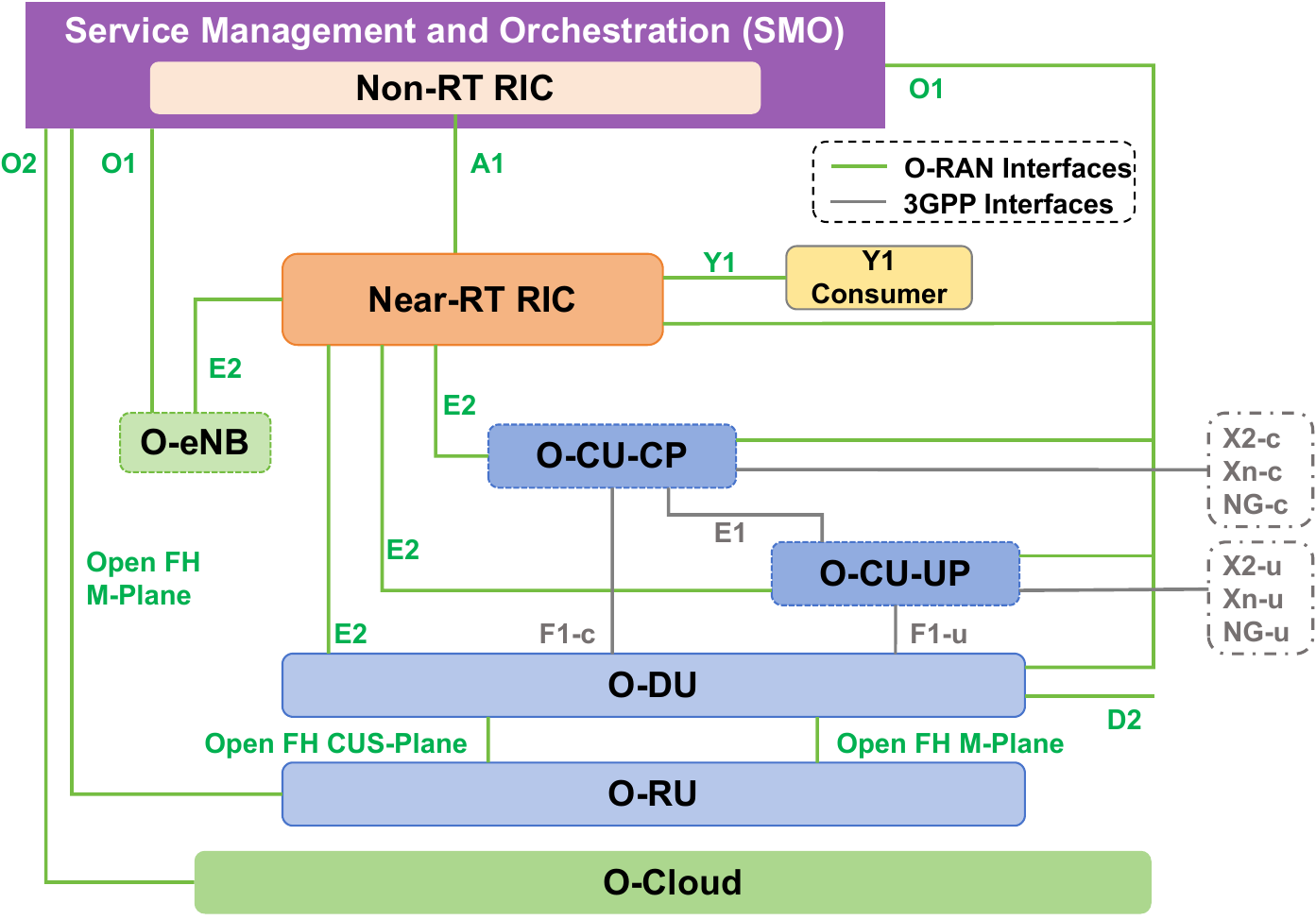}
	\caption{The latest ORAN  Architecture by O-RAN Alliance \cite{oran_architecture_2025}.}
	\label{Latest_ORAN_A}
\end{figure}

The latest ORAN architecture defined by the O-RAN Alliance is illustrated in Fig.~\ref{Latest_ORAN_A}. We offer a brief introduction about each component as follows:
\begin{itemize}
    \item \textbf{SMO}: SMO provides the end-to-end (E2E) lifecycle management, such as the orchestration and fault, configuration, accounting, performance, and security (FCAPS) operations, of the ORAN network functions and RICs \cite{baranda2025architectural}.
    \item \textbf{O-eNB}: O-eNB may take the form of either a legacy eNB or an ng-eNB, since the ORAN architecture also supports the integration of LTE base-stations.

    \item \textbf{O-Cloud}: O-Cloud serves as a cloud computing platform to host the relevant ORAN network functions, supporting software components, and management and orchestration functions. 

    \item \textbf{Y1 Consumer}: Y1 Consumer refers to the entity within or outside of the public land mobile network (PLMN) trust domain that consumes the Y1 services produced by the near-real-time RIC (Near-RT RIC) \cite{oran_y1_interface_2024}.
\end{itemize}
ORAN architecture defines disaggregated functional entities, including the ORAN radio unit (O-RU), ORAN distributed unit (O-DU), and ORAN centralized unit (O-CU), which are interconnected through standardized open interfaces specified by the O-RAN Alliance and 3GPP. The O-RAN Alliance and
3GPP has specified different split options as discussed later in Section \ref{ORAN_NTN_proposed}. Following the typical functional split Option 7.2x \cite{larsen2018survey, oran_architecture_2025}, the main functionalities of these entities are:
\begin{itemize}
    \item \textbf{O-RU}: O-RU typically hosts the lower physical layer (L-PHY), and supports radio frequency (RF) processing operations. 
    \item \textbf{O-DU}: O-DU usually hosts the medium access control (MAC), radio link control (RLC) protocols, and higher physical layer (H-PHY) functionalities, such as data segmentation, scheduling, multiplexing, and other baseband processing.
    \item \textbf{O-CU}: O-CU usually hosts the higher-layer protocols such as the radio resource control (RRC), packet data convergence protocol (PDCP), service data adaptation protocol (SDAP), and also supports functionalities like network slicing, connection, and mobility management. O-CU can be further divided into control plane (O-CU-CP) and user plane (O-CU-UP), with O-CU-CP implementing PDCP and RRC, and O-CU-UP implementing PDCP and SDAP. 
\end{itemize}
In addition, two types of RICs, including non-real-time RIC (Non-RT RIC) and Near-RT RIC, enable the intelligent RAN optimization across different layers and timescales \cite{oranwhitepaper-1,oran_architecture_2025}. Their key functionalities are summarized as follows:
\begin{itemize}
    \item \textbf{Near-RT RIC}: Near-RT RIC is a software platform that hosts use-case specific applications known as xApps, and performs the control and optimization over the RAN elements and resources in near-real-time, usually between 10 ms and 1 s. xApps are microservice-based applications tailored to operate on the Near-RT RIC, which could be provided by third-party software companies. Some examples of xApp include mobility management, traffic steering, load balancing, and admission control.
    \item \textbf{Non-RT RIC}: Non-RT RIC is placed within the SMO framework and hosts applications named as rApps to support intelligent RAN optimization radio resource management (RRM) in non-real-time, usually longer than 1 s. rApps are modular third-party applications tailored for the Non-RT RIC, which are designed to run on any vendor’s Non-RT RIC. Non-RT RIC usually performs the long-term data analytics and decisions based on the data from Near-RT RIC, and also provides guidance, data enrichment, and management of AI/ML models for Near-RT RIC via the A1 interface.
\end{itemize}
To provide essential background on the interfaces defined by 3GPP but adopted by the O-RAN Alliance \cite{oran_architecture_2025,ORAN-survey-2025}, brief descriptions are presented as follows:
\begin{itemize}
    \item \textbf{E1}: E1 interface connects the O-CU-UP and O-CU-CP entities and enables their efficient coordination within O-CU. 
    \item \textbf{F1}: F1 interface connects the O-CU and O-DU, and also consists of the F1-c and F1-u interfaces for control and user plane information, respectively.
    \item \textbf{NG}: NG interface connects the O-CU to the 5G core (5GC), with NG-c connecting the O-CU-CP with the access and mobility function (AMF), and NG-u connecting the O-CU-UP to the user plane function (UPF).
    \item \textbf{X2}: X2 interface is adopted by ORAN to support interoperability profile specifications. X2 connects the O-CU with other eNBs in an E-UTRAN NR dual connectivity (EN-DC) configuration, with X2-c and X2-u for transmitting the control and user plane information.
    \item \textbf{Xn}: Xn interface is adopted by ORAN to support interoperability profile specifications. Xn connects the O-CU with other gNBs, with Xn-c and Xn-u for transmitting the control and user plane information.
\end{itemize}
Furthermore, we introduce the new interfaces defined by the O-RAN Alliance below \cite{openRAN-4, oran_architecture_2025}.
\begin{itemize}
    \item \textbf{A1}: A1 interface connects the Non-RT RIC in SMO and the Near-RT RIC, and supports three types of services: policy management, enrichment information, and ML model management.
    \item \textbf{O1}: O1 interface connects the SMO and Non-RT RIC with the ORAN managed elements, such as the O-CU, O-DU, Near-RT RIC, and O-eNB. O1 interface is defined to support the SMO's management of the ORAN  network functions.
    \item \textbf{O2}: O2 interface is used by SMO for O-Cloud management, orchestration, and workflow management. This interface provides two types of services: O2 deployment management services and O2 infrastructure management services.
    \item \textbf{E2}: E2 interface is a control plane interface connecting the Near-RT RIC with an E2 Node, such as O-CU, O-DU, or O-eNB. E2 interface allows the Near-RT RIC to control RRM and other functionalities of the E2 nodes.
    \item \textbf{Open FH}: Open FH interface connects the O-DU and O-RU, and includes the control user synchronization (CUS) plane and management (M) plane. In hybrid mode, the Open FH M-Plane interface connects the O-RU to the SMO for FCAPS functionality.
    \item \textbf{Y1}: Y1 interface enables the Y1 consumers to subscribe or request the RAN analytics information provided by Near-RT RIC \cite{oran_y1_interface_2024}. 
    \item \textbf{D2}: D2 interface is newly defined as the interface between two O-DU connected to the same O-CU-CP, and supports NR carrier aggregation of carriers configured in different O-DU nodes \cite{oran_D2_interface_2025}.
\end{itemize}

\subsubsection{KEY FEATURES}
The ORAN  architecture exhibits several fundamental features that distinguish it from conventional RAN systems, which can be summarized as: 
\begin{itemize}
    \item \textbf{Disaggregation}: The conventional BS functionalities are decoupled into three logical units, O-RU, O-DU, and O-CU. The O-CU can be further divided into O-CU-UP and O-CU-CP. This disaggregated architecture enables the operators to update or upgrade the hardware and software independently. 

    \item \textbf{Open Interfaces}:
    The open interface design enables full interoperability and flexible reconfiguration of RAN components, supporting an open and competitive ecosystem. This openness facilitates seamless integration of multi-vendor equipment, promotes innovation in network deployment, and supports the delivery of diverse and customizable user services.

    \item \textbf{Virtualization}: The virtualization or softwarization of RAN resources decouples software functions from dedicated hardware, enabling flexible deployment on general-purpose virtualized computing platforms. This approach can support efficient resource utilization, enhance scalability, and allow dynamic adaptation to varying network demands compared with traditional vendor-locked architectures.

    \item \textbf{Intelligence}: The ORAN architecture incorporates two RICs to enable intelligent network optimization. AI /ML algorithms will be more efficiently employed to determine and implement policies and control actions at different timescales.
\end{itemize}

\subsection{NTN}
In this subsection, we review the basic background on NTN, including its components, payload types, and characteristics. 
A broad range of survey and magazine articles has examined NTNs from different perspectives. Several works focus on 3GPP standardization and evolution from 5G to 6G, including surveys \cite{el2023introduction,azari2022evolution,rinaldi2020non}, as well as magazine articles \cite{hosseinian2021review,lin20215g,saad2024non,wigard2023ubiquitous}. Other works emphasize applications and system architectures, including surveys \cite{openRAN-NTN-1, he2024non} and magazine papers \cite{jamshed2025non}. In addition, several recent works, including \cite{deng2025distributed,kaleem2024emerging,mahboob2024revolutionizing}, investigate emerging technologies and intelligent NTN paradigms, with particular attention to AI-based solutions such as generative AI and reinforcement learning. These works complement the scope of this section and provide readers with a comprehensive overview of NTN research, which covers both technological evolution and standardization progress.

\begin{figure}[ht]
  \centering
  \includegraphics[width = 8.5cm]{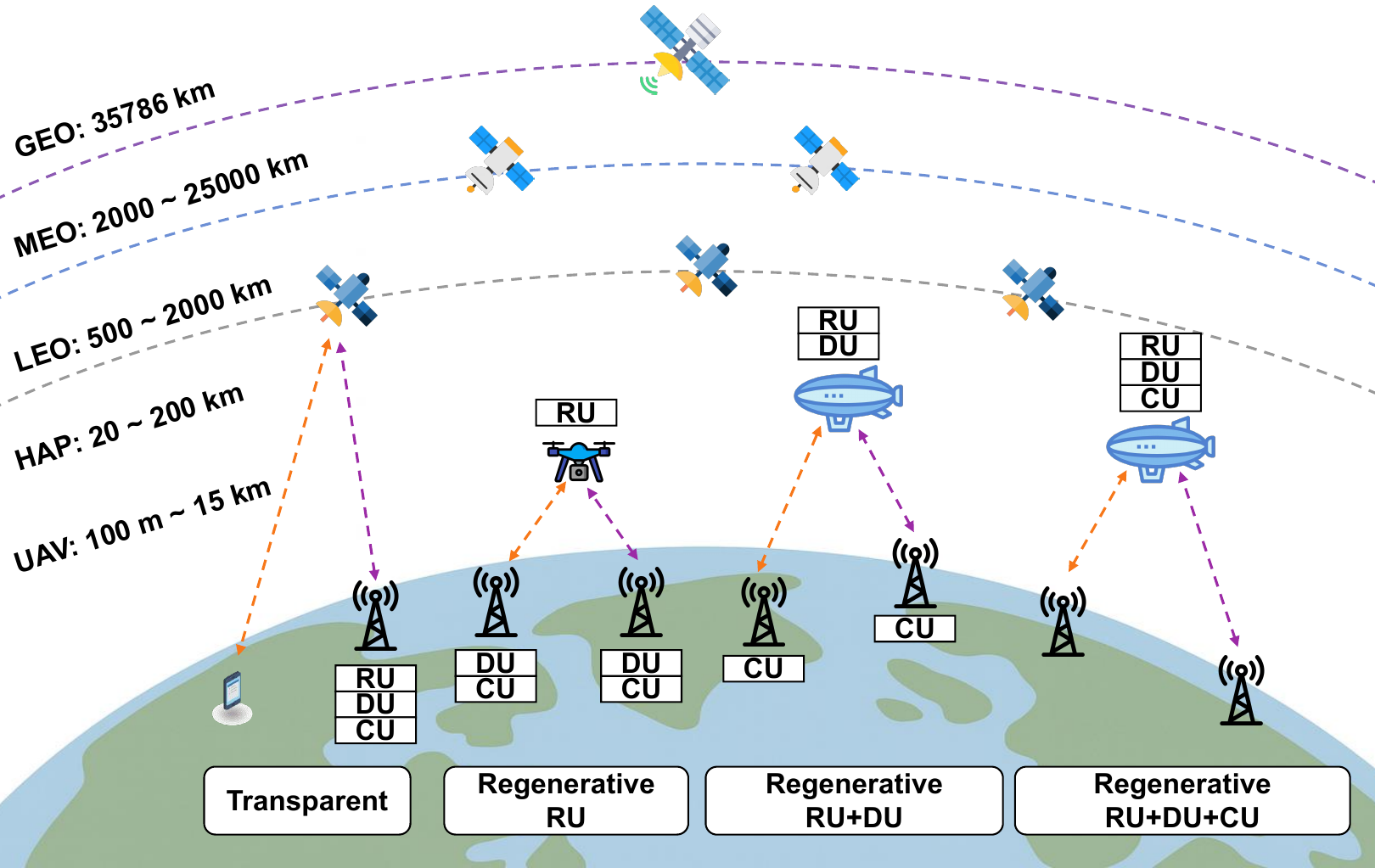}
  \caption{NTN components and payload options.}
  \label{fig:ntn_payload}
\end{figure}

\subsubsection{COMPONENTS}
In the NTN architecture, the NTN platforms carry the payload between user equipment and ground stations via service and feeder links. Generally, NTN involves different space-borne or airborne platforms, which can be classified into five categories based on their operating orbit altitudes, as shown in Fig. \ref{fig:ntn_payload}.

\begin{itemize}
    \item \textbf{Geostationary Earth Orbit (GEO) satellites}: GEO satellites typically maintain a fixed position at an orbit altitude of 35,786 km above the Earth's equator. They provide extremely wide coverage and maintain fixed visibility from the ground, but they suffer from high propagation delay and significant path loss. GEO systems are commonly used for broadcasting, satellite television, and large-scale communication services that are less sensitive to latency \cite{mahboob2025transforming}.
    \item \textbf{Medium Earth Orbit (MEO) satellites}: MEO satellites are located at an orbit altitude of 2,000 to 25,000 km above the Earth. They offer a balance between coverage area, latency, and constellation size, achieving lower delays compared with GEO while requiring fewer satellites than Low Earth Orbit (LEO) satellite constellations to achieve global coverage \cite{al2022survey}. MEO satellites are widely used in navigation and positioning systems such as Global Positioning System (GPS), China's BeiDou navigation satellite system (BDS), Russia's Global Navigation Satellite System (GLONASS), and the European Union's Galileo \cite{jiao2025geostationary}.
    \item \textbf{Low Earth Orbit (LEO) satellites}: LEO satellites operate at altitudes of approximately 500 to 2,000 km above the Earth. This orbit altitude allows them to achieve a much shorter propagation delay and reduced path loss compared with a satellite system with a higher altitude. However, LEO satellites face several challenges, including their relatively smaller coverage, the need for frequent handovers, and the strong Doppler effect because of the high mobility \cite{deng2025orthogonality}. In practice, typical LEO constellations such as Starlink have demonstrated real-world use in delivering global mobile broadband internet and improving connectivity in remote regions \cite{hui2025review}.
    \item \textbf{High Altitude Platforms (HAPs)}: The HAPs, such as airplanes, balloons, and airships, usually reach the stratosphere region with an altitude ranging from 20 km to 50 km \cite{shang2024enhancing,openRAN-NTN-1}. They combine the advantages of satellites and terrestrial systems by maintaining a quasi-stationary state. Meanwhile, they offer relatively wide coverage, low latency, and flexible deployment. These characteristics make them well suited for civilian applications, such as providing broadband internet access to rural areas \cite{liu2023joint}. 
    \item \textbf{Uncrewed Aerial Vehicles (UAVs)}: UAVs typically operate at low altitudes ranging from 0.1 km to 15 km. Due to their high mobility and deployment flexibility, they are suitable for rapidly providing emergency coverage in post-disaster scenarios \cite{kaleem2024emerging}. However, UAVs suffer from limited endurance, rapidly varying communication channels, and complex interference management. Beyond emergency response, UAVs are also expected to play a central role in the emerging low-altitude economy (LAE), enabling services such as aerial logistics, infrastructure monitoring, and precision agriculture \cite{ahmed2025toward}.
\end{itemize}

\subsubsection{PAYLOAD OPTIONS}
There are typically two main types of payload architecture defined for NTN by 3GPP: transparent and regenerative \cite{rinaldi2020non, mahboob2025transforming}. The detailed schemes behind these two architectures are introduced as follows: 

\textbf{Transparent payload:}
With transparent payload architecture, the NTN platform simply relays the signal from users to the terrestrial gNodeB (gNB), which consists of CU, DU, and RU \cite{wigard2023ubiquitous,saad2024non}.

\textbf{Regenerative payload:}
The regenerative payload architecture typically can be classified into three categories depending on the configuration of the gNB components on the NTN platform.
\begin{itemize}
    \item Only RU: the NTN platform is equipped solely with the RU, responsible for basic RF functions or partial physical layer processing, depending on the selected functional split and NTN's capabilities. DU and CU are deployed on the ground, resulting in long propagation delays on the wireless fronthaul link to RU \cite{shah2025urllc}.

    \item Partial gNB (RU+DU): the NTN platform can take on some of the processing tasks that are normally performed by the full gNB on the ground. Then the signal will be further processed by the gNB-CU on the ground for further processing. This setting can improve the performance of the network, especially in areas with high traffic demand \cite{rihan2023ran}.
    
    \item Full gNB (RU+DU+CU): the NTN platform would have complete processing capabilities to regenerate the signal in either the feeder link or the service link, which allows for better performance, such as lower latency and higher data rates \cite{mahboob2025transforming}. 
\end{itemize}

\subsection{AI-NATIVE FOR COMMUNICATION}\label{subsec_ai_native}
In recent years, extensive research efforts have been devoted to exploring technologies beyond 5G toward the realization of 6G networks. As the 6G network evolves into a programmable and flexible cloud-native implementation, the integration of AI/ML techniques will be the key feature of the 6G network. In this subsection, we introduce the basic concepts and the architecture of the AI-Native for communication. 

\begin{figure*}[h!]
	\centering
    \includegraphics[width=0.83\textwidth]{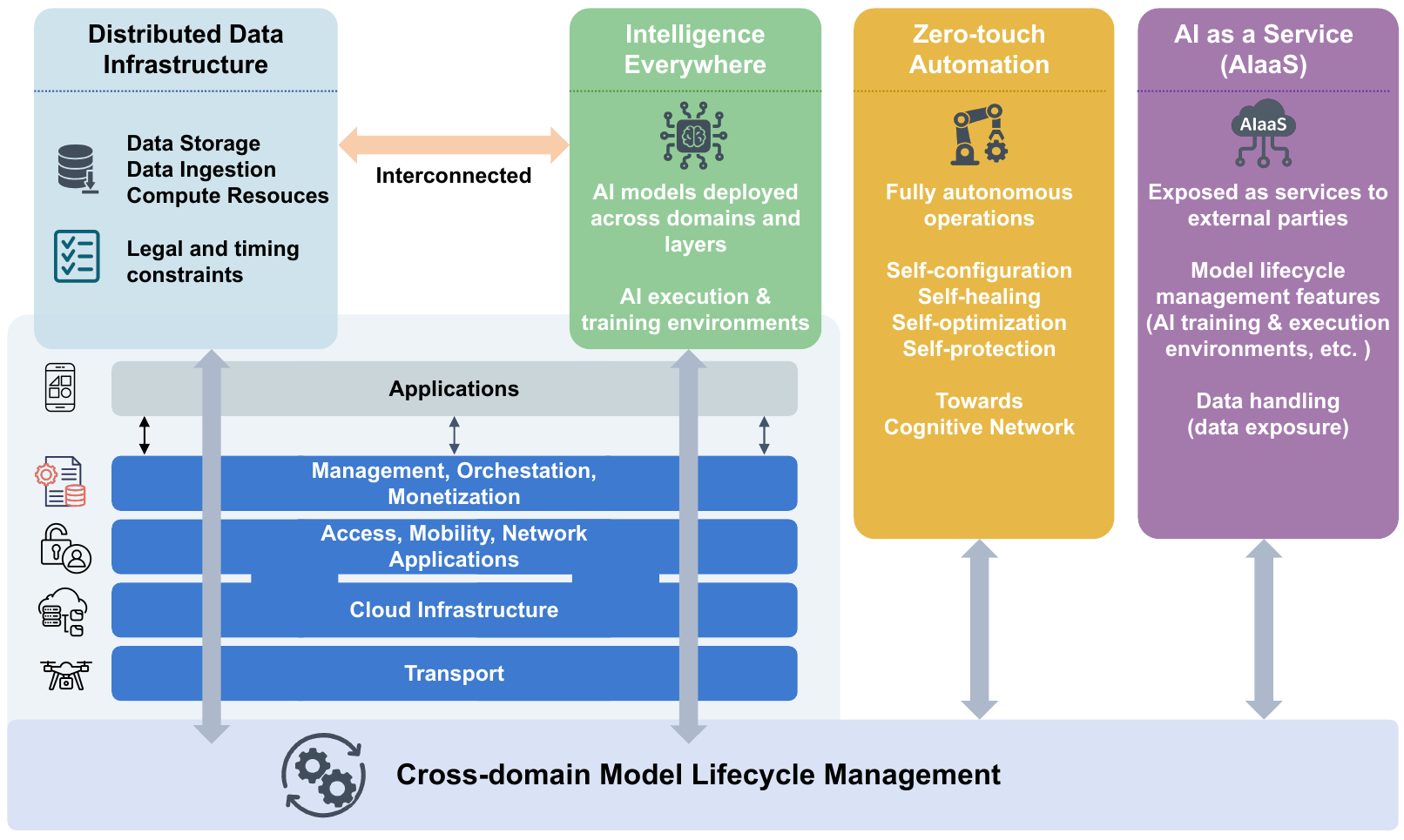}
	\caption{AI-Native architecture based on \cite{iovene2023defining}.}
	\label{Fig_ai_native}
\end{figure*}

\subsubsection{CONCEPT}
Typically, the AI-based communication can be divided into two stages, i.e., AI-Aided and AI-Native \cite{baek2023ai}. Specifically, an AI-Aided method aims to optimize specific communication modules, which has been shown to be practical and feasible in numerous studies \cite{deng2025distributed, AI-for-TN-NTN-1, kaleem2024emerging,khan2023ai}. An AI-Native method will enable the system to have intrinsic trustworthy AI capabilities, where AI is naturally embedded across system design, deployment, operation, and maintenance. In addition, AI-native implementations leverage a data-driven and knowledge-based ecosystem that continuously consumes and produces information. This ecosystem can help enable new AI functionalities, or replace the static and rule-based mechanisms with adaptive and learning-based intelligence when needed \cite{iovene2023defining}.

Several recent studies and reports have discussed the concept of AI-Native communications and its applications in ORAN and NTNs \cite{3GPP_AI_ML_NG_RAN,baek2023ai,khan2023ai,iovene2023defining}. The 3GPP report \cite{3GPP_AI_ML_NG_RAN} discussed the integration of AI/ML in NG-RAN and 5G-Advanced systems as a step toward 6G evolution. The authors in \cite{baek2023ai} outlined key challenges and future directions for realizing AI-Native communication systems. In \cite{khan2023ai}, the authors analyzed AI-driven RAN design for 6G, highlighting major research issues, challenges, and potential solutions. Additionally, the Ericsson white paper \cite{iovene2023defining} defined AI-Native as a key enabler for intelligent telecommunication networks and presented an architectural framework and maturity model.

\subsubsection{ARCHITECTURE}
Ericsson has presented a concise and generic AI-Native architecture in \cite{iovene2023defining}. Building upon this concept, we extend and refine its essential aspects to develop a detailed version, as illustrated in Fig.~\ref{Fig_ai_native}.

As shown in the figure, the architecture is structured around several core layers, including transport, cloud infrastructure, access and mobility, management and orchestration, and applications. These layers interact closely with distributed intelligence and data resources to support seamless execution and adaptation of AI models across different domains. Moreover, the cross-domain model lifecycle management helps enable coordinated and trusted intelligence, constantly improving and following data changes, to achieve system-wide E2E performance gains.

The AI-Native architecture is characterized by four fundamental aspects \cite{iovene2023defining}:
\begin{itemize}
    \item \textbf{Intelligence Everywhere}: AI workloads can be deployed and executed wherever needed, across every layer, domain, and physical location from cloud to edge. Meanwhile, AI execution environments should be available everywhere, while the AI training environments might also be co-located when needed.
    
    \item \textbf{Distributed Data Infrastructure}: This aspect enables the data and knowledge sharing across different layers. It supports the aspect of intelligence everywhere by ensuring the availability of data and computing resources. In addition, it also interacts with the model orchestrator for the transportation of data or intelligence. For example, under stringent requirements on the timing of the data, it is often more efficient to move intelligence closer to the data source, rather than transporting data to intelligence. 
    
    \item \textbf{Zero-touch Automation}: The AI-Native architecture aims for autonomous management of AI and data processes. This zero-touch automation aspect consists of self-configuration, self-healing, self-optimization, and self-protection capabilities, thereby facilitating the realization of a cognitive network.

    \item \textbf{AI as a Service (AIaaS)}: Certain functionalities, such as AI model training or execution environments, and data handling, can be exposed as services to external parties. This concept, known as AIaaS, transforms networks into open innovation platforms, where service providers and even end users can flexibly access and utilize AI resources on demand. 

\end{itemize}

\section{NTN DEVOPS CHALLENGES AND SOLUTIONS}
\label{NTN_devops_challenges}

\begin{figure}[h!]
	\centering
\includegraphics[width = 8.5cm]{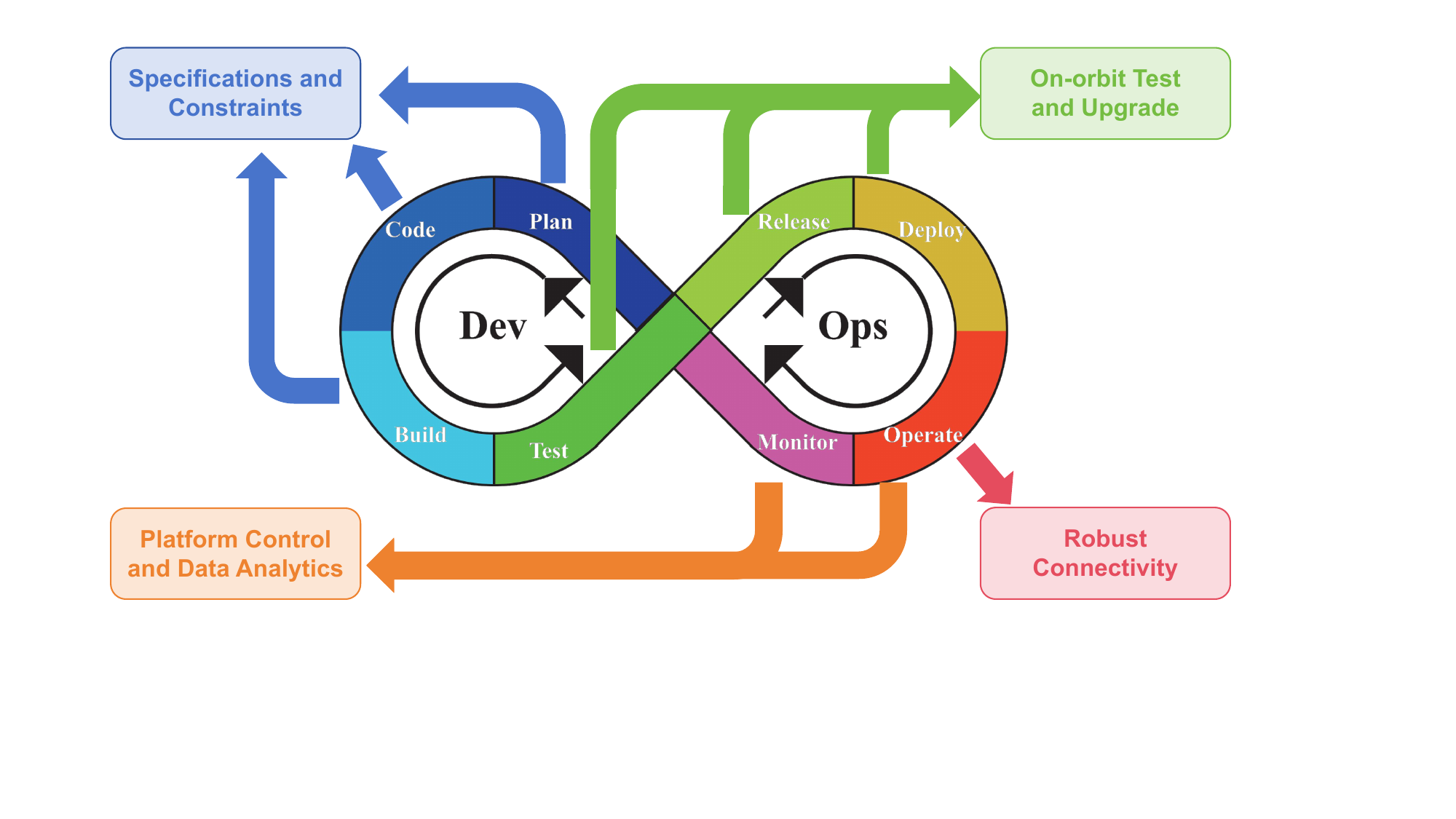}
	\caption{DevOps lifecycle and corresponding challenges in NTN.}
	\label{DevOps_lifecycle}
\end{figure}

Different from traditional TNs, the DevOps lifecycle in NTN faces multiple unique challenges \cite{herrera2025tutorial,Corici2023}, as illustrated in Fig.~\ref{DevOps_lifecycle}. In this section, we focus on analyzing these DevOps challenges of NTN and the corresponding solutions via integrating ORAN principles. 

\begin{table*}[!htb]
    \caption{Typical specifications of NTN platforms \cite{deng2025distributed}.}
    \begin{center}
        \begin{tabular}{|c|c|c|c|c|c|c|}
            \hline
            \rowcolor[HTML]{E0E0E0}
            \textbf{Specification}  &\textbf{Untethered UAV}&\textbf{Tethered UAV}&\textbf{HAP}&\textbf{LEO}&\textbf{MEO}&\textbf{GEO}\\
            \hline			
            \hline
            \cellcolor[HTML]{E0E0E0}\textbf{Loading Capability} &2.7 kg&15 kg& 140 kg&-&-&-\\
            \hline
            \cellcolor[HTML]{E0E0E0}\textbf{Power Supply} &0.3 kWh&30 kWh&20 kWh&-&-&-\\
            \hline
            \cellcolor[HTML]{E0E0E0}\textbf{Latency}  &1 ms&1 ms&1 ms&30-50 ms&150 ms&600 ms\\
            \hline
            \cellcolor[HTML]{E0E0E0}\textbf{Endurance} &30 mins&24 hours& 2 months&5-7 years&10-15 years& 15-20 years\\
            \hline
            \cellcolor[HTML]{E0E0E0}\textbf{Deployment Cost}&Low&Low& Medium&High&High&High\\
            \hline
            \cellcolor[HTML]{E0E0E0}\textbf{Network Type}  &1 Micro/Pico &1 Macro &7 Multiple Macro &61 Multiple Macro &61 Multiple Macro &61 Multiple Macro \\
            \hline
        \end{tabular}
    \label{table_ntn_specification_summarization}
    \end{center}
\end{table*}

\subsection{NTN SPECIFICATIONS AND CONSTRAINTS}
The typical weights of macro massive MIMO RU and baseband unit (BBU) are over 10Kg and 5Kg, respectively \cite{Ericsson_MIMO}. Noticeably, in the 5G era, the maximum power consumption of a 64T64R active antenna unit (AAU) is 1000–1400 W, and that of a BBU is about 2000 W \cite{Huawei_5Gpower}. However, different NTN platforms have various specifications and constraints, which should be considered in the NTN development stage \cite{Haq_2025,shang2025novel}. 

The DJI flagship untethered UAV platform Matrice 350 RTK achieves a loading capability of 2.7 Kg with 31 minutes of flight time, which cannot carry the macro BS equipment discussed above \cite{DJI_350}. Although DJI has customized a delivery UAV platform with a 40 Kg loading capability (i.e., DJI Flycart30). The flight time is limited to 8 minutes, which fails to guarantee continuous connections to the users \cite{DJI_flycart30}. In addition, the battery is originally designed to support the UAV flight with a typical power supply of 0.3 kWh (e.g., TB65 from DJI). Apart from untethered UAVs, tethered UAVs have also been regarded as an emerging NTN platform due to their tethered systems. The DG-M20 tethered UAV, from Dagong Technology Co., Ltd, can carry 15 Kg payloads hovering at 100 m for 24 hours \cite{DG_tetheredUAV}. Furthermore, the HAP has also attracted attention from both academia and industry due to its relatively lower altitude and quasi-stationary position. Taking the HAP from Stratospheric Platforms Ltd as an example, with a wingspan of 60 m, it can support a maximum of 140 Kg payload and can provide a 20 kWh power supply \cite{Strato_HAP}. The typical specifications of NTN platforms are summarized in Table \ref{table_ntn_specification_summarization}, including loading capability, power supply, and other related aspects.

To overcome the above specification limitations, the standardized function splits and open interfaces defined in ORAN can be leveraged. These mechanisms provide flexibility for NTN MNOs to customize onboard components O-RU/O-DU/O-CU from different vendors, thereby meeting the specification requirements of NTN platforms.
For example, the commercial untethered UAV can be utilized to implement a micro or pico network with integrated O-RU/O-DU/O-CU and 5GC on board, which acts as an E2E flexible deployment option and avoids adding peripheral equipment for FH, midhaul, and backhaul transmissions. On the contrary, the power cable and optical fiber in the tethered UAV system allow it to serve as a macro network with various onboard options, including O-RU, O-RU+O-DU, and  O-RU+O-DU+O-CU, which can be configured according to the latency requirement of the service.

\subsection{NTN ROBUST CONNECTIVITY} 
Due to the mobility of the NTN platforms and the varying channel conditions, one important challenge is to guarantee robust connectivity during operation, especially for the LEO satellites with high speeds. To solve this challenge, a simple solution is to deploy a massive constellation of satellites, which can provide redundant access links and transportation links under a highly dynamic network topology. For example, SpaceX from the USA has successfully launched more than 5000 Starlink satellites into orbit. In addition, China has proposed GW-A59 and GW-2 sub-constellations as GuoWang (national network) mega-constellation, a large satellite internet project aimed at providing broadband services. It is worth noting that GW-A59, with 6,080 satellites, is intended to orbit at a lower altitude (below 500 km), while GW-2, with 6,912 satellites, will orbit at a higher altitude (1,145 km above Earth). However, the massive deployment of satellites imposes a heavy financial investment burden on a single NTN MNO and requires long development periods.

The integration of ORAN's interoperability into NTN has the potential to bring new opportunities to solve the challenge above by enabling network sharing among MNOs. At the end of 2024, 3GPP identified Indirect Network Sharing (INS) as the next evolution step towards 5G-Advanced Rel-20, which aims to enhance network co-construction and sharing capability \cite{3gpp23501}. Specifically, INS was regarded as an effective solution to mitigate the compatibility issues between the network elements of sharing parties, e.g., shared satellite access network and two or more core networks of terrestrial participating operators via multiple operator core network (MOCN) \cite{3gpp32130}. However, to avoid direct connections between network elements of different MNOs, INS is based on the build of a hosting network in advance by all participating MNOs, which cannot adapt to the high mobility and complex NTN environments. ORAN surpasses the INS concept by enabling direct interconnections among MNOs, and has the potential to achieve real-time on-demand sharing based on the instantaneous network status, thereby offering greater flexibility in NTN.
For clarity, consider a scenario in which O-RU, O-DU, and O-CU components of MNO A experience failures due to space weather events. In this case, MNO A can temporarily access the O-RU, O-DU, and O-CU resources of MNO B through the E2, E1, F1, and Open FH interfaces to restore service continuity. Meanwhile, to strengthen the interface security during on-demand sharing among MNOs, O-RAN Alliance has also defined dedicated security mechanisms for each interface \cite{oran_security_2025}. Furthermore, the virtualization of ORAN enables the deployment of Docker technology, which can also enhance data isolation and security during data processing among MNOs.

\subsection{NTN PLATFORM CONTROL AND DATA ANALYTICS}
In traditional TNs, the BS is usually installed on fixed terrestrial infrastructure with a sufficient and stable power supply, such as cell towers, where the network performance is independent of the terrestrial infrastructure during operation. However, in the NTN scenario, the instantaneous position, velocity, and battery status of the platforms impact the network performance significantly during handover, beam management, or power allocation. To improve the network performance, it is important to design an interface between the NTN platform and onboard communication components. This interface can support dynamic optimization of communication algorithms by adapting to instantaneous platform states, and also enable proactive platform control. However, following the vendor-locked principle in traditional RAN, the vendor is required to develop both the NTN onboard components and the NTN platform to ensure compatibility, which may increase development complexity and investment costs.

The recent O-RAN Alliance white paper in \cite{ORAN_white_paper} has proposed integrating ORAN into NTNs to address these challenges by providing standardized and open interfaces between the RIC/SMO and NTN platform. However, it mainly focuses on offloading the additional terrestrial traffic, instead of designing the open interfaces between the onboard component and the NTN platform for real-time control and optimization. Therefore, further standardization of open interfaces between the NTN platform and O-RU/O-DU/O-CU is essential to enhance network performance and deployment flexibility.
For example, DJI has provided standard application programming interface (API) functions to control UAV mobility and access real-time UAV status. Specifically, the Roll, Pitch, and Yaw information and battery status can be accessed periodically under pre-defined binary control and command (C\&C) formats. In addition, the design of such interfaces also requires close collaboration between NTN platform manufacturing communities and the O-RAN Alliance to accurately capture the diverse characteristics of NTN platforms.

\subsection{NTN ON-ORBIT TEST AND UPGRADE} 
Traditional RAN is mainly based on vendor-locked hardware, and most network functions and algorithms are implemented on specialized hardware with limited upgrade capability, such as application specific integrated circuit (ASIC), field-programmable gate array (FPGA), and digital signal processor (DSP). The specialized hardware is effective in TN due to the easy access and low upgrade costs for testing and upgrading. However, in the NTN scenario, launching NTN platforms for the testing or deployment of new features is much more costly and time-consuming \cite{sandri2023ns,liu2024challenges}. Therefore, it is essential to achieve on-orbit testing and upgrading to improve flexibility and shorten DevOps cycles in NTNs. In this regard, based on network function virtualization (NFV), ORAN enables the algorithms to be deployed on commercial off-the-shelf (COTS) hardware, such as central processing unit (CPU) or graphics processing unit (GPU). This virtualization capability will allow MNOs to test and upgrade the related components in NTN in a more agile DevOps manner.

\section{AI-NATIVE ORAN-BASED NTN FRAMEWORK}
\label{ORAN_NTN_proposed}
Motivated by the advantages of ORAN-based NTN and the envisioned AI-based 6G networks, in this section, we propose an AI-Native ORAN-based NTN framework. The architecture details and key technological enablers of the proposed framework are presented below.

\begin{figure*}[h!]
	\centering
    \includegraphics[width=\textwidth]{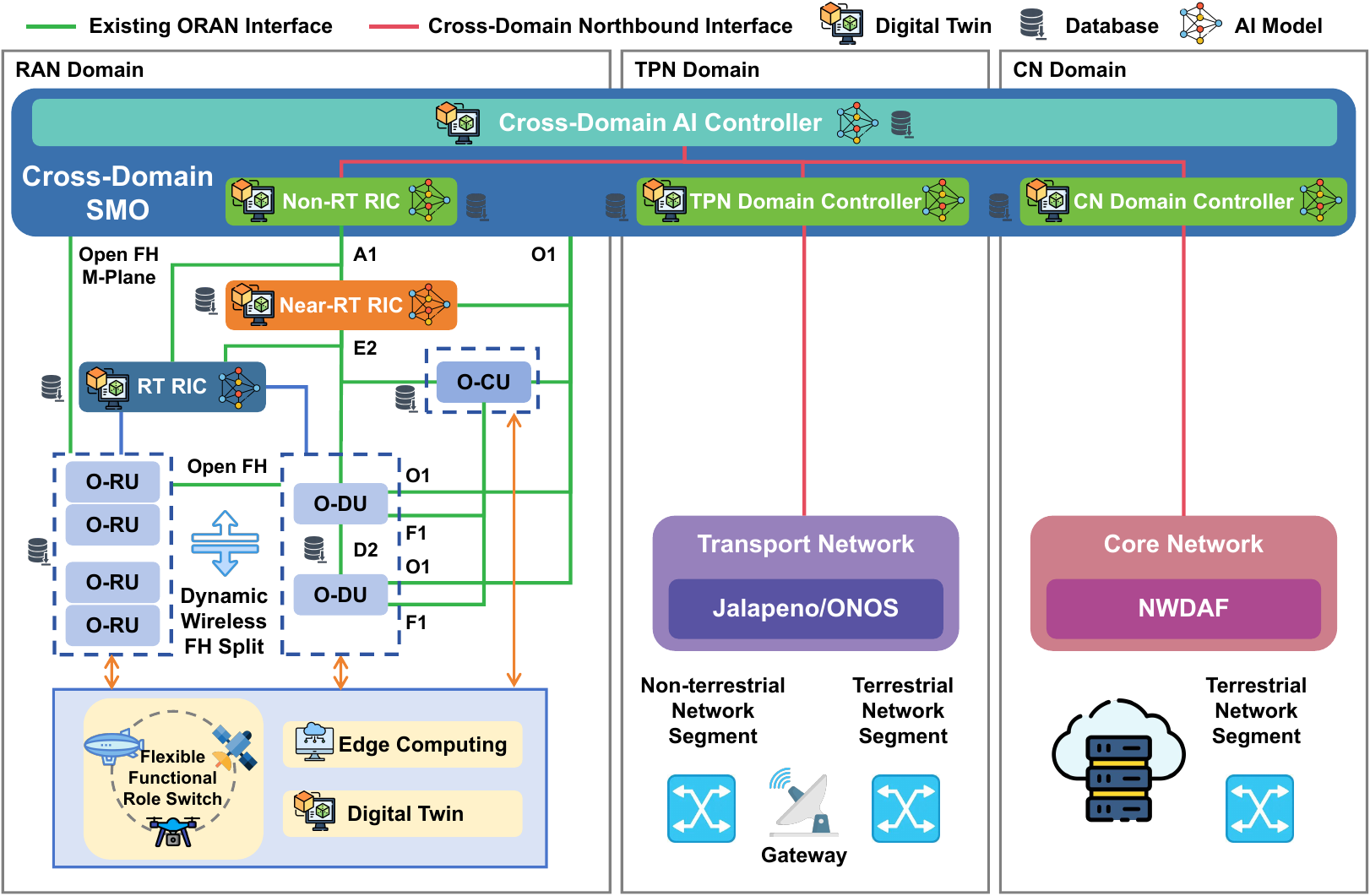}
	\caption{Proposed orchestrated AI-Native ORAN-based NTN framework.}
	\label{Proposed_ORAN_NTN}
\end{figure*}

\subsection{ARCHITECTURE}
The proposed framework is illustrated in Fig.~\ref{Proposed_ORAN_NTN}, which spans the RAN, transport network (TPN), and core network (CN) domains.  Specifically, the RAN domain is implemented within the NTN segment, while the TPN domain includes both NTN and TN segments, and the CN domain is deployed in the TN segment. Each domain is managed and orchestrated by a controller, which is equipped with AI models and digital twin technology to enhance data analysis and predictive capabilities. 

The cross-domain SMO includes three domain controllers and a cross-domain AI controller. It is also responsible for O-Cloud management and orchestration, and RAN network function operation and maintenance as defined by the O-RAN Alliance. The domain controllers, including Non-RT RIC, TPN domain controller, and CN domain controller, host AI models to infer the optimized variables in RAN, TPN, and CN domains, respectively. Each domain controller can access the domain-specific data for AI model training and execution. Due to the hierarchical timescale characteristic, the TPN domain controller can access historical data from the RAN domain, and the CN domain controller can further utilize historical data from both the RAN and TPN domains.

Beyond single-domain intelligence, a cross-domain AI controller is introduced at the cross-domain SMO layer for AI lifecycle management. This controller can operate in two modes. For the first mode, the cross-domain AI controller serves as a central server, which helps facilitate the coordination of MADRL or aggregate the local models for a global model in FL. For the second mode, the cross-domain AI controller aims to achieve E2E optimization, and it hosts an AI model to infer the optimal policy based on the three single-domain controllers, thereby guaranteeing service-level agreements (SLAs).

It is worth noting that the proposed framework is designed to be an AI-Native system by supporting each AI-Native aspect defined in Section \ref{subsec_ai_native}, and the details of the realization of each AI-Native aspect are presented in Table \ref{tab_mapping_ai_native}. Furthermore, the proposed framework consists of four key technological enablers, including dynamic wireless FH split, multi-tier RIC structure, flexible functional role switch, and cross-domain SMO. Each key enabler helps facilitate the implementation and operation of this proposed framework, which will be discussed in detail in the following subsections.

\begin{table*}[!t]
\centering

\caption{Detailed realization of AI-Native aspects in the proposed ORAN-based NTN Framework}
\renewcommand{\arraystretch}{1.25}
\begin{tabular}{>{\centering\arraybackslash}m{3.2cm} |m{13.5cm}}
\toprule
\textbf{AI-Native Aspect} & \textbf{Realization in the Proposed ORAN-based NTN Framework} \\
\midrule

\textbf{Intelligence Everywhere}
&
\begin{itemize}
    \item Multi-tier deployment of AI models across Non-RT RIC, Near-RT RIC, and RT RIC to support decision-making at different timescales.
    \item Cross-domain AI controller supports coordinated intelligence across RAN, TPN, and CN domains.
\end{itemize}
\\
\midrule

\textbf{Distributed Data Infrastructure}
&
\begin{itemize}
    \item Supports cross-layer and cross-domain data and knowledge sharing through open interfaces of ORAN and cross-domain SMO.
    \item Provides distributed computing resources by enabling edge computing on NTN platforms and cloud computing in the cross-domain SMO or Non-RT RIC.
\end{itemize}
\\
\midrule

\textbf{Zero-touch Automation}
&
\begin{itemize}
    \item AI-assisted FCAPS operation in cross-domain SMO enables the self-configuration, self-healing, self-optimization, and self-protection.
    \item xApps and rApps form closed-loop control mechanisms for near-real-time and non-real-time self-optimization.
    \item The dynamic wireless FH split and flexible functional role switch enables the self-configuration to adapt to rapidly changing NTN scenarios.
\end{itemize}
\\
\midrule

\textbf{AI as a Service (AIaaS)}
&
\begin{itemize}
    \item Cross-domain AI controller enables shared access to AI-based inference services across RAN, TPN, and CN.
    \item The AI lifecycle management and computing resource of this framework can be utilized by third parties via open interfaces for their AI model training or execution.
\end{itemize}
\\

\bottomrule
\end{tabular}
\label{tab_mapping_ai_native}
\end{table*}

\subsection{ENABLER-1: DYNAMIC WIRELESS FH SPLIT}
\begin{figure}[h!]
  \centering
  \includegraphics[width = 8.4cm]{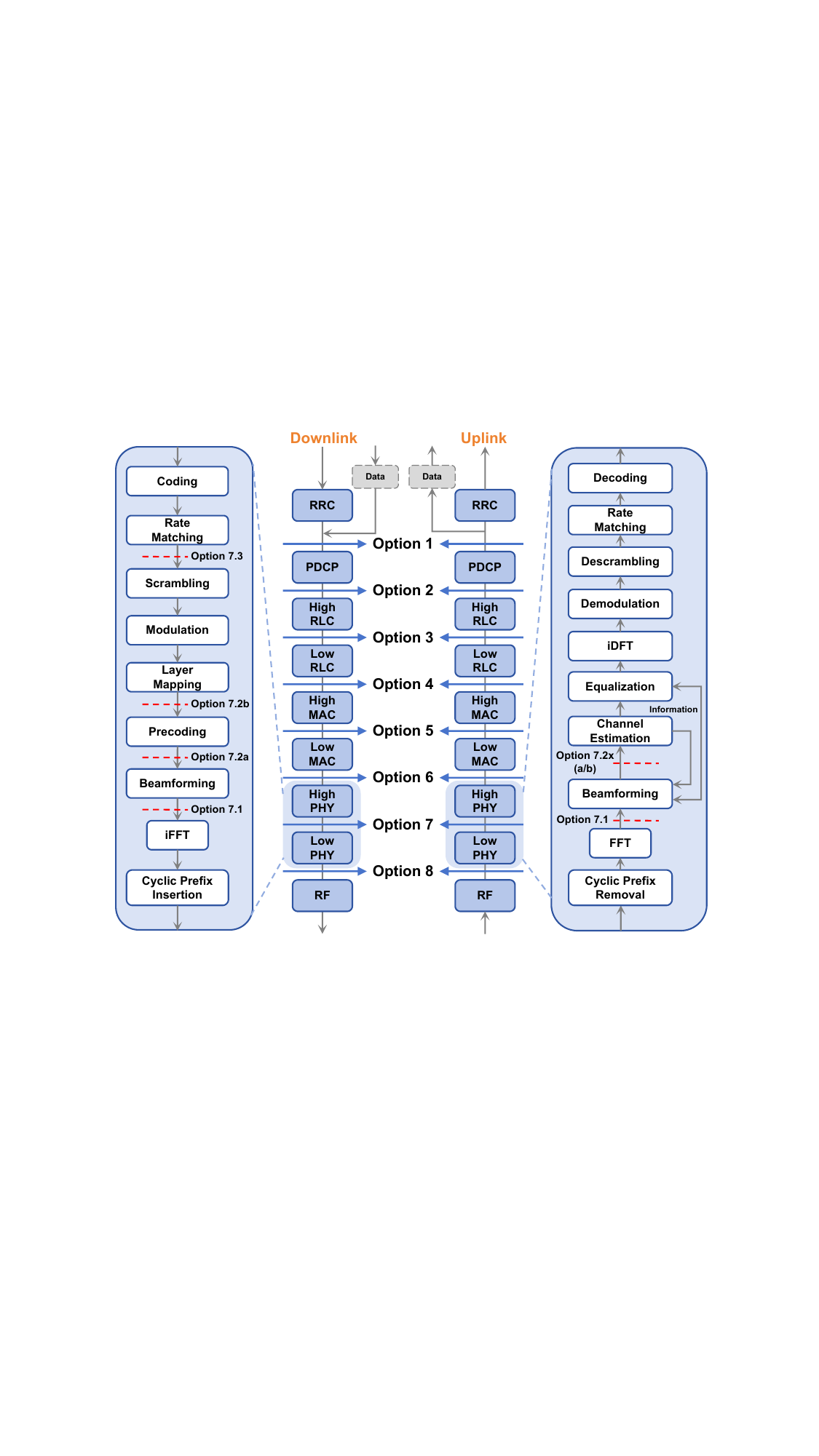}
  \caption{Functional split options defined by 3GPP.}
  \label{fig:functional_split}
\end{figure}

The gNB’s operation is conceptually structured as a series of functions, with the distribution of these functions defined by functional splits. Within the 3GPP framework, eight distinct functional split options exist, as shown in Fig. \ref{fig:functional_split}. Each split option describes how the logical nodes interrelate to one another and what specific activities each undertakes \cite{openRAN-8}. From Option 1 to 8, there is a higher degree of centralization and a notable improvement in resource management efficiency, which leads to reduced O-RU complexity and associated costs. However, this comes at the expense of imposing high data rates and strict latency requirements on the FH between O-DU and O-RU \cite{larsen2018survey}. 

Typically, the O-RAN Alliance lends its support to the Option 7.2x variant for networks with high-capacity and high-reliability requirements, and this variant further splits the functions in the L-PHY based on the PHY split in Option 7. This option is made possible by using the eCPRI interface for FH. In Option 7.2x, the O-CU/O-DU manages RRC, PDCP, RLC, MAC, H-PHY, and parts of L-PHY, while the O-RU handles RF and the rest of L-PHY. The 'x' in this split name can be either 'a' or 'b', depending on whether some of the PHY functionalities reside in H-PHY or L-PHY \cite{openRAN-7}. 
Without loss of generality, our following discussion is based on the downlink transmission function split. As shown in Fig. \ref{fig:functional_split}, for Option 7.2a, the PHY functions belonging to O-RU are beamforming, inverse fast Fourier transform (iFFT), and cyclic prefix insertion. The rest of the PHY functions are in the O-DU. In Option 7.2b, the precoding function is added to the O-RU. As the FFT function is deployed in the O-RU, the FH data rate and required bandwidth are significantly reduced compared to Option 8. The Option 7.2x enables a relatively simple RU whose size and power consumption support network densification and enable sharing by multiple operators, which is better suited for macrocell deployment in urban areas in terrestrial networks \cite{andersson2021functional}.

The different FH split option choices impact the network via three aspects, including the data rate requirement, the latency requirement, and the cooperation scale. Each aspect is examined in detail as follows.

\subsubsection{Data Rate}
The required FH data rate increases significantly when the functional split occurs in lower PHY layers, due to the need to transmit raw In-phase and Quadrature (IQ) samples or partially processed signals \cite{larsen2018survey}. In addition, the data rate is also influenced by the number of bits required to represent matrix elements in the precoding and beamforming functions, as it may vary depending on the dynamic range and precision requirements given the underlying matrix multiplication operations in lower PHY layers. Therefore, both FH split options and the quantization effect should be considered to satisfy the FH data rate constraint \cite{park2014fronthaul}.

The FH data rate constraint in NTN is more stringent than in traditional TN, where the optical fiber is typically deployed between O-RU and O-DU. However, in NTNs, O-RU and O-DU may be deployed on different platforms depending on the payload capabilities, making it challenging to employ wireline-based FH solutions. For example, the authors in \cite{lee2025ran} proposed a LEO-based NTN architecture with the O-CU located on the ground, while the RU and DU were deployed on separate satellites. The FH transmission was achieved via inter-satellite links (ISLs) by optimizing signal compression. In addition, the authors in \cite{zhang2021optimizing} considered the free-space optical (FSO) communication for the FH link due to its high data rate and bandwidth. Nevertheless, FSO-based FH remains challenged by atmospheric turbulence and the beam alignment problem, particularly for mobile NTN platforms such as HAPs or UAVs operating in more complicated atmospheric conditions. Therefore, to guarantee the stringent FH data rate constraint, it is essential to carefully design the functional split option and compression techniques tailored to the specific characteristics of different NTN platforms.

\subsubsection{Latency}
Functional split at lower PHY layer imposes stricter latency requirements on FH transmissions due to data exchange between O-RU and O-DU. This becomes particularly critical for uplink transmissions under Option 7.2a/7.2b, where the beamforming function resides in the O-RU, while channel estimation and equalization are performed in the O-DU. This split creates a latency-sensitive control loop among these three functions, as shown in the uplink part of Fig. \ref{fig:functional_split}. Specifically, the beamforming operation at the O-RU may rely on outdated channel state information (CSI) due to the high latency of transferring CSI back and forth between O-RU and O-DU, which further results in a trade-off between user throughput and FH data rate \cite{ericsson2023massive}.

However, in NTNs, the FH propagation delay increases significantly due to the long separation distance between O-RU and O-DU \cite{larsen2018survey}. In \cite{seeram2024feasibility}, the authors analyzed the maximum propagation distance between O-RU and O-DU when deployed on different NTN platforms. Specifically, the ideal latency requirement for lower PHY layer FH split options should be within 0.25 ms. In \cite{veisi2025non}, the authors evaluated the one-way FH latency between O-RU and O-DU, reporting values ranging from 2.1 to 18 ms when deployed on different NTN platforms. 
Although the FH latency requirement can be relaxed if disabling features such as CSI reporting or reciprocity-based channel estimation in NTN, meeting FH latency constraints while maintaining these features to enhance data rates remains an important challenge.

\subsubsection{Cooperation Scale}
Multiple FH split options at the PHY layer facilitate coordination among multiple O-RUs to boost performance via distributed MIMO (DMIMO) or coordinated scheduling. For example, considering an ORAN-based NTN architecture with distributed O-RUs and one centralized O-DU, Option 7.2x split has the potential to enable centralized cooperation subject to stringent FH data rate and latency requirements. On the other hand, Option 2, with relaxed requirements on FH data rate or latency, results in limited cooperation capability since each O-RU implements the functions of the PHY and MAC layers locally. Therefore, the trade-off between FH requirements and cooperation among O-RUs should be investigated in the FH split design.

Unlike TN, the wireless FH links in NTNs make it more challenging to achieve efficient centralized cooperation. In \cite{Abdelsadek2022}, the authors identified synchronization, outdated CSI, and signalling overhead issues in proposed LEO-based DMIMO networks, where each LEO satellite is assumed to connect to a central unit through ISL. In \cite{Abdelsadek2023}, the authors proposed to provide connectivity for handheld devices via LEO DMIMO networks with perfect FH links. Furthermore, \cite{Riera-Palou2022} discussed both centralized and distributed precoding strategies for LEO cell-free MIMO networks, which require various levels of CSI information sharing. Despite these advances, determining the optimal wireless FH split configuration that enables efficient cooperative processing among NTN platforms still remains an open research problem.\\

To solve the challenges above, we propose a \textbf{\textit{dynamic wireless FH split}} scheme that leverages NFV capabilities in ORAN-based NTN to address the FH split challenges. Existing works related to dynamic FH split mainly focused on the data rate and latency requirements in TNs \cite{Vajd2022,Datsika2021,Diez2021,Villegas2024,pousa2024road,Girycki2024}. Building on these works, we extend the dynamic FH split solution for NTN, as shown in Fig.~\ref{Proposed_ORAN_NTN}. Specifically, the cross-domain SMO supports both hierarchical and hybrid modes for O-RU and O-DU FH split reconfiguration. In the hierarchical mode, the cross-domain SMO transmits FCAPS information over O1 to the O-DU, which uses the Open FH M-Plane interface to manage the O-RU. In the hybrid mode, the cross-domain SMO provides FCAPS information over the Open FH M-Plane interface directly to the O-RU, or over O1 to the O-DU, and then uses the Open FH M-Plane interface to manage the O-RU.  

To optimize the dynamic wireless FH split performance, the hierarchical deep reinforcement learning (DRL) integrated with the graph neural network (GNN) can be employed as a promising method. For instance, within the NTN segment, a typical RAN domain can be naturally modeled as a graph structure with O-RU, O-DU, and O-CU as nodes, where the FH and midhaul links serve as edges between O-RU and O-DU nodes, and between O-DU and O-CU nodes, respectively. Within the DRL framework, the nodes and edges features of the graph are treated as the system state, the reward is defined according to the optimization objective function, such as throughput or latency, and the action is the FH split option for each O-RU and O-DU pair.
It is worth noting that dynamic wireless FH split optimization is a delay-tolerant service with a relatively long update period. As a result, the Non-RT RIC can utilize the O1 and Open FH M-plane interfaces to collect the global network information for graph construction. To guarantee the sum capacity of FH links does not exceed the midhaul capacity of an O-DU and also meet latency requirements, constrained DRL approaches based on Lagrangian methods and action masking techniques can be employed. Based on the FH split solution, the hierarchical DRL can be utilized to further determine the cooperation scale among O-RUs, aiming to enhance the quality of service for cell-edge users.

\subsection{ENABLER-2: MULTI‑TIER RIC STRUCTURE}

As discussed above, the O-RAN Alliance has introduced Near-RT and Non-RT RICs to manage network control and optimization across different timescales and application scenarios \cite{openRAN-4}. Using the Non-RT control loop, the Non-RT RIC provides guidance, enrichment information, and management of ML models for the Near-RT RIC. To enhance the performance of RICs in ORAN-based NTN, it is essential to consider the impact of non-negligible latency and the system challenges when facilitating the distributed learning frameworks.

\subsubsection{Latency}
In the NTN scenario, the Near-RT RIC and O-CU/O-DU/O-RU may be deployed separately on different NTN platforms, leading to non-negligible propagation delays. The delay from Near-RT RIC to O-DU and from O-DU to O-RU remains an important challenge, which will significantly degrade the intelligent Near-RT optimization performance.
Meanwhile, considering the split Option 7.2x, the scheduling and beamforming functions in O-DU/O-RU are typically required to be executed in a real-time timescale ($<\!10\! $ ms), while the modulation function in O-RU needs to be executed in a smaller timescale ($<\!1\!$ ms) \cite{openRAN-4}. This fact makes it challenging to continue employing Near-RT RIC for optimizing such functions with a small timescale.

\subsubsection{Distributed Learning}
The introduction of the Near-RT RIC in ORAN facilitates the deployment of distributed learning frameworks such as federated learning (FL) and multi-agent deep reinforcement learning (MADRL). In this context, Near-RT RIC serves as the central server to coordinate the training phase among O-RUs and O-DUs. 
Specifically, the FL framework aims to reduce communication overheads and preserve data privacy. In each FL communication round, O-RUs/O-DUs upload gradients obtained from local data samples to the Near-RT RIC, which aggregates the gradients and updates the global model using a gradient descent algorithm. The updated global model is then broadcast back to all participating O-RUs and O-DUs \cite{farajzadeh2025federated}.
Furthermore, the MADRL algorithms extend traditional DRL to enable the coordination among multiple agents. The centralized training and distributed execution (CTDE) has been regarded as a promising solution to achieve the trade-off between performance and communication overheads \cite{deng2025distributed}. Specifically, within the CTDE framework, each O-RU or O-DU shares information with the Near-RT RIC during the training phase to enhance the coordination and cooperation capability. After convergence, each O-RU and O-DU can perform the action inference based on local observation during execution. 

However, in the AI-Native ORAN-based NTN framework, ensuring robust and continuously learning AI models, such as MADRL and FL, inevitably requires addressing several inherent system challenges. These challenges include highly heterogeneous transmission latencies, intermittent connectivity, and non-independent and identically distributed data across different NTN orbits. For example, the transmission latencies during multiple rounds of information exchange will slow down the convergence of distributed learning. Meanwhile, heterogeneous latencies among RICs and O-CU/O-DU/O-RU entities may lead to asynchronous updates, thereby disrupting the continuity of AI models.

These challenges highlight the need for a dynamic and platform-aware RIC placement strategy that considers the involved O-DUs/O-RUs, their channel conditions, and each NTN platform's available computational resources. This observation further reveals the edge–cloud trade-off that must be carefully considered in multi-tier RIC deployment for NTNs. To be more specific, deploying RICs at NTN may help reduce control-loop latency, alleviate feeder-link and backhaul loads, and keep detailed KPMs local, but it is constrained by the limited computing and energy resources on the NTN platform. In contrast, deploying RICs at the cloud or ground can help provide sufficient processing capability and the global information, but the time-critical control loops may be degraded due to round-trip delays.\\

To solve the challenges above, a practical solution is the adoption of a \textit{\textbf{multi-tier RIC structure}}, as illustrated in Fig.~\ref{Proposed_ORAN_NTN}. In this architecture, a real-time RIC (RT-RIC) can be implemented within the O-RU or O-DU to control functions, such as scheduling and beamforming ($<\!10\! $ ms), or the modulation function in even shorter timescales ($<\!1\! $ ms) \cite{openRAN-TN-1, NTN-challenges-2}. The Near-RT RIC is mainly responsible for supporting time-sensitive xApp inference and short control loops, such as RAN slicing or handover, on a near-real-time timescale ($10\,\text{ms}-1\,\text{s}$).
The computationally intensive tasks, including heavy AI model training, policy optimization, and cross-constellation coordination, are delegated to the Non-RT RIC ($>\!1\!$ s) deployed on the ground. Concretely, UAV platforms with strict SWaP constraints can offload computationally intensive AI tasks to ground controllers, whereas HAPs/LEO will host more complex Near-RT functionalities locally and perform model or policy updates less frequently. This multi-tier RIC structure ensures the AI models and data sources are accessible across different layers or timescales for training and execution, which further facilitates the zero-touch automation.

\subsection{ENABLER-3: FLEXIBLE FUNCTIONAL ROLE SWITCH}
It is worth noting that the high mobility of NTNs can easily lead to link failure. Meanwhile, since the ORAN typically involves multiple functional entities, a failure in any entity may disrupt the connection, particularly when the malfunction occurs in higher-level components such as the O-CU. However, in the existing ORAN architecture, the functional roles of the O-CU, O-DU, and O-RU across different network platforms are typically fixed after the initial configuration. This static configuration limits the system’s resilience in adapting to dynamic network conditions, further limiting the scalability of NTN systems.

To address the challenges above, a promising direction is to design a novel framework that enables the \textbf{\textit{flexible functional role switch}}, which allows the NTN platforms to switch functional roles to O-RU/O-DU/O-CU based on real-time network conditions and functional demands, as illustrated in Fig.~\ref{Proposed_ORAN_NTN}. This approach is implemented based on NFV and the use of COTS components, which can effectively enhance the scalability and resilience of NTNs \cite{pousa2024road,hojeij2023dynamic}. 
The role switch can typically be triggered by fault-related events or on-demand transition events. Specifically, the fault-related events include O-CU/O-DU failures, excessive FH latency, congestion, insufficient power supply, etc. In the proposed framework, the cross-domain SMO periodically collects telemetry data related to infrastructure, platform status, and ORAN system operation through the O1 and O2 interfaces \cite{ls2025falcon}. By aggregating and analyzing the telemetry data using AI/ML models such as digital twin \cite{mukherjee2023open}, these events can be detected or even predicted in advance. For the on-demand transition event, when the communication system transitions to serve a different scenario, such as emergency communication under an earthquake or a tsunami, the role switch may also be triggered to provide specific service and maintain service continuity.

\begin{figure}[h!]
	\centering
\includegraphics[width=0.45\textwidth]{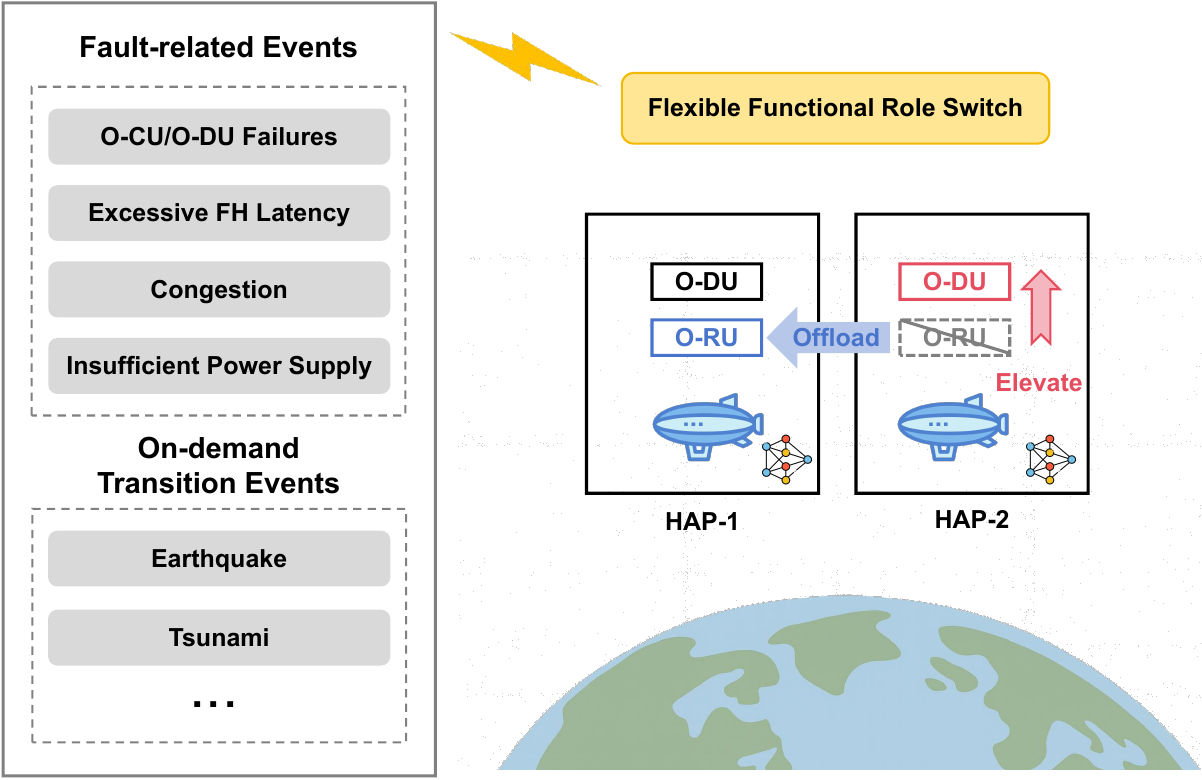}
	\caption{Process of flexible functional role switch.}
	\label{flexible_role_switch}
\end{figure}

To better illustrate the function of role switch, consider a scenario in which a HAP initially operates as an O-RU, as shown in Fig. \ref{flexible_role_switch}. When triggered by either fault-related events or mission-transition events, the HAP can dynamically elevate its functional role to an O-DU or even an O-CU by downloading or upgrading the related software. In the meantime, its original O-RU role can be intelligently offloaded to a neighboring HAP through network reconfiguration mechanisms, ensuring the completeness of ORAN components and service continuity.

The role switch enhances network adaptability and latency reduction, especially in dynamic or post-disaster environments where traditional terrestrial networks may be compromised. However, implementing the role switch involves addressing challenges such as computational constraints and power supply to support the tasks executed by the new functional role. For example, the computational loads and the power consumption for O-CU are typically much heavier than those of a single O-RU. Meanwhile, to ensure a seamless functional role switch, different NTN platforms need to maintain efficient cooperation and synchronous role configuration. Moreover, to prevent unauthorized role elevation or offloading, the security issues of role switch need to be addressed through enhanced authentication techniques and continuous cross-domain SMO monitoring. After the functional migration, data exchanges in backhaul, midhaul, or FH will also require real-time routing optimization to deal with the dynamic NTN topology.

\subsection{ENABLER-4: CROSS-DOMAIN SMO}
Cross-domain SMO plays a key role in ORAN-based NTNs by enabling network planning, dynamic and efficient resource allocation, proactive management, and prediction of complicated topologies in NTNs. Despite its importance, current 3GPP or O-RAN Alliance standards have not yet defined or implemented a cross-domain SMO structure. Specifically, 3GPP primarily focuses on the CN domain through the network data analytics functions (NWDAF), while the O-RAN Alliance focuses on the RAN domain via RICs. Although 3GPP has introduced the concept of cross-domain management data analytics (MDA), it mainly addresses the ML model lifecycle and the processing of management and network data, which is mainly utilized for network-level and delay-tolerant optimization, instead of E2E intelligent orchestration \cite{3gpp28105}.

To address this gap, this paper proposes a \textbf{\textit{cross-domain SMO}} layer, as illustrated in Fig. \ref{Proposed_ORAN_NTN}, to enable E2E intelligent orchestration across RAN, TPN, and CN domains. Specifically, the cross-domain SMO relies on multiple data analytics functions (DAFs) operating across different domains. The representative examples include the NWDAF, Jalapeno, and the open network operating system (ONOS). In addition, it also relies on the RICs of the ORAN-based NTNs through the appropriate northbound interfaces \cite{ORAN-management-1}.  It is worth noting that this orchestration can be achieved through an intelligent or E2E software layer. This layer will support seamless integration and unified management of air interface schemes, ORAN-based NTN configurations, and the updates of hardware and firmware.

The integration of cross-domain SMO layer can streamline and accelerate the deployment of ORAN-based NTNs for both system integrators and service providers. Moreover, it enables more efficient network operation by reducing both CapEx and OpEx. This reduction is achieved by minimizing the number of isolated operational support systems (OSSs), network management systems (NMSs), and operational tools from different vendors across various domains. To illustrate the functionality of the proposed cross-domain SMO, we take the cross-domain slice adaptation as a representative example. To enhance the slice adaptation, a typical control loop spanning the RAN, TPN, and CN can be implemented based on a hierarchical MADRL framework. Specifically, the Non-RT RIC optimizes user-level resource allocation within each slice, the TPN domain controller optimizes the path selection of TPN for each slice, and the CN domain controller optimizes computing resource allocation for different slice categories. Through coordinated decision making across domains, E2E slice adaptation can be achieved to satisfy heterogeneous quality of service (QoS) requirements \cite{LIAO2025102821}.

\section{USE CASES} \label{use cases}
\begin{figure}[h!]
	\centering
    \includegraphics[width=0.45\textwidth]{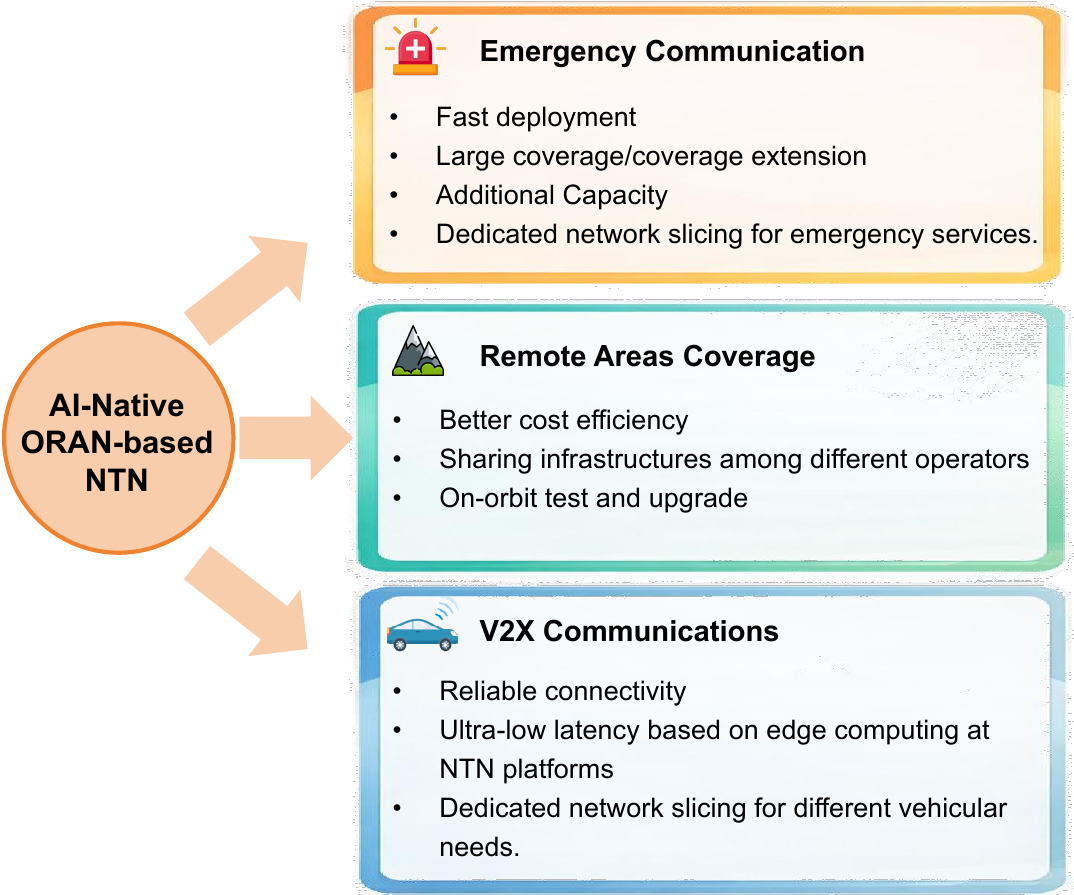}
	\caption{Use cases of the AI-Native ORAN-based NTN framework.}
	\label{use_cases}
\end{figure}

The utilization of ORAN-based solutions for NTNs will enable a variety of innovative use cases in the upcoming 6G networks. In this section, we present three representative use cases of the proposed AI-Native ORAN-based NTN framework, including emergency communication, remote areas coverage, and V2X communications. The realization of these use cases is fundamentally supported by the proposed framework through fast deployment, better cost efficiency, and reliable connectivity, as shown in Fig. \ref{use_cases}.

\subsection{EMERGENCY COMMUNICATION}  
Emergency communication usually requires the fast deployment of networks to replace destroyed terrestrial infrastructure and provide additional capacity for post-disaster recovery. 
Although the satellites are promising for such scenarios due to their wide coverage, the fixed orbit limits their ability to offer timely connectivity. Satellite orbit transfer techniques can be used to adjust satellite orbits by configuring orbital parameters such as altitude, inclination, and eccentricity. However, for a single MNO, establishing E2E links via the Hohmann transfer orbit may take several days to complete. 
To overcome these limitations, the proposed ORAN-based NTN framework supports interoperation among MNOs, allowing multiple MNOs to coordinate and share the access, fronthaul, midhaul, and backhaul link resources. Through the multi-MNO collaboration, an emergency communication network can be established more rapidly, without relying solely on slowly adjusting the satellites' orbit from a single MNO. 
Furthermore, since ORAN can support network slicing, the proposed framework can create dedicated and prioritized network slices for emergency services. This ensures that critical communications, such as voice, data, drone video feeds, and sensor information, are allocated sufficient bandwidth even under network congestion. In addition, the RICs can dynamically allocate resources to further prioritize emergency communication traffic, thereby supporting timely and effective crisis management.

\subsection{REMOTE AREAS COVERAGE} 
Although NTNs offer the potential for ubiquitous connectivity, the high costs associated with testing, deployment, and operation may discourage a single MNO from investing in sufficient NTN infrastructure for remote areas coverage. For example, in LEO satellite-based NTNs, a single MNO would need to deploy hundreds to thousands of satellites to build its own network, with each satellite serving a limited geographic area over a limited time window. 
Moreover, to reduce the substantial deployment cost of base stations in NTN, MNOs across different regions can benefit from sharing the NTN infrastructure. This can be realized through the standardization of interfaces between base stations and NTN platforms, as well as among NTN platforms operated by different MNOs. This approach follows the same principle as O-RU and O-DU sharing in ORAN. Based on the on-orbit test and upgrade, along with the collaboration of multiple MNOs, the proposed ORAN-based NTN framework has the potential to significantly reduce the costs of remote areas coverage.

\subsection{V2X COMMUNICATIONS}
In B5G and 6G networks, V2X communications are expected to advance substantially, enabling applications such as autonomous driving, cooperative awareness, and real-time traffic coordination. The 5G Automotive Association (5GAA) has emphasized the potential of NTNs for connected vehicles, with early deployments anticipated by 2027 and broader adoption by 2030 \cite{5gaa2024ntn}. By combining the virtualization and intelligence of ORAN with the flexibility and wide coverage of NTN, the proposed ORAN-based NTN framework can effectively address the existing key challenges in V2X communication. Specifically, the ORAN-based NTN framework can provide reliable connectivity in underserved remote areas, which makes it also suitable for supporting remote driving functions, such as safety alerts and collision avoidance \cite{oran2025private}. Moreover, the ultra-low latency can be supported for safety-critical V2X communication by leveraging edge computing at NTN platforms. The AI-enabled RICs in the proposed framework help dynamically optimize vehicular communication, prioritize traffic during congestion or emergencies, and adapt to vehicle mobility in real time. Additionally, ORAN's programmable network slicing can extend over NTN links to deliver dedicated slices for different vehicular needs, ensure URLLC for autonomous control, and support enhanced mobile broadband (eMBB) slices for diagnostics \cite{alam2024slicing}. This capability to tailor services across hybrid terrestrial and non-terrestrial networks will be important for supporting future heterogeneous V2X demands.

\section{FUTURE RESEARCH DIRECTIONS}
\label{future directions}
\begin{figure}[h!]
	\centering
\includegraphics[width=0.45\textwidth]{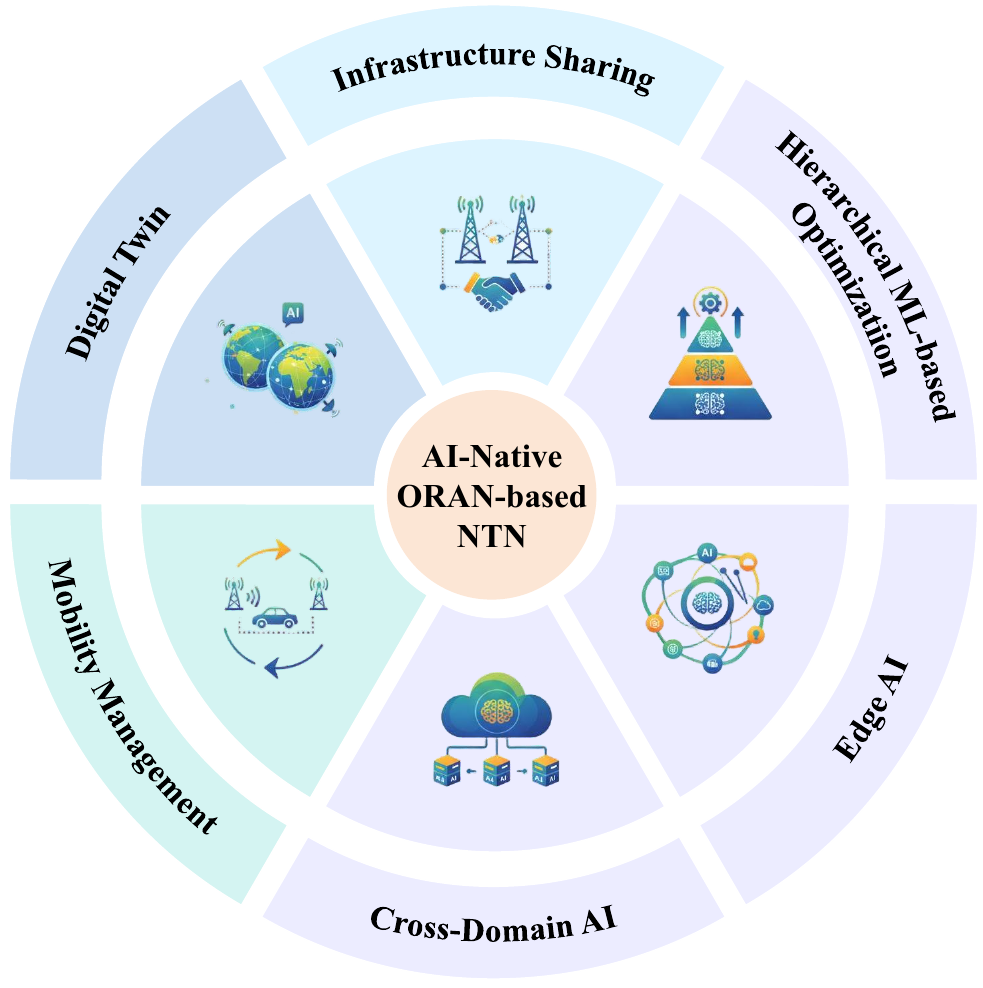}
	\caption{Future research directions for AI-Native ORAN-based NTN framework.}
	\label{Future_direction}
\end{figure}

In this section, we further conduct an in-depth analysis of the prospective future research directions for the proposed framework, which is shown in Fig.~\ref{Future_direction}. These directions encompass both business and technical aspects of AI-Native ORAN-based NTNs. Specifically, the business perspective focuses on infrastructure sharing, which enables cost-efficient and collaborative network deployment. From the technical perspective, several promising AI/ML methods are identified, including hierarchical ML-based optimization, edge AI, and cross-domain AI. Furthermore, at the system level, mobility management and digital twin are investigated, serving as essential mechanisms for effective network control and accurate predictive system modeling.

\subsection{INFRASTRUCTURE SHARING} 
\subsubsection{DEFINITION}
Infrastructure sharing among MNOs is a key business model in the industry, with competitors turning into partners to minimize the investment costs and expand the coverage of MNOs \cite{GSMA2025}. In a traditional TN, infrastructure sharing can be broadly classified into two categories: passive and active. Passive sharing involves the sharing of telecommunication sites, towers, buildings, and facilities such as power supplies and air conditioning. Active sharing, on the other hand, involves the components of the active network layer, such as the antennas, base stations, or even core network elements \cite{ITU2008_sharingInfrastructure}. Due to the increasing development and deployment costs in NTN, the infrastructure sharing in the NTN environment remains an urgent challenge to be solved.
\subsubsection{RECENT ADVANCES AND PROMISING DIRECTIONS}
To the best of our knowledge, few works have explicitly investigated or systematically analyzed NTN-based infrastructure sharing strategies \cite{lee2025ran, he2025accountability}. In \cite{lee2025ran}, the authors proposed a joint power allocation and FH compression optimization in ORAN-based NTN for multiple MNOs sharing the LEO satellites, but it lacks policy-level mechanisms for sharing resources, such as spectrum, beams, and computing resources, across multiple operators or verticals in NTN environments. In \cite{he2025accountability}, the authors analyzed several ORAN use cases related to collaboration and resource sharing among multiple parties, with little attention to NTN integration. Therefore, we provide a brief discussion on NTN-based infrastructure sharing under the proposed ORAN-based NTN framework, aiming to offer preliminary insights and serve as a reference for future studies in this area.

In NTN-based infrastructure sharing, the active sharing will involve the sharing of the RAN, TPN, and CN, while passive sharing will also include the sharing of orbits or NTN platforms. Based on the proposed framework, the active sharing for multiple NTN MNOs can be achieved based on the interoperability and NFV of ORAN. The interoperability allows each MNO to utilize other MNOs’ DUs or CUs to establish an E2E communication link under a dynamic network topology. NFV enables virtualized DU and CU pools, which can support the resource sharing among O-RUs and O-DUs of multiple MNOs \cite{he2025accountability}. However, unlike terrestrial passive sharing, NTN-based passive sharing depends on the mobility and coverage characteristics of different NTN platforms and faces several unique challenges. Specifically, UAVs and HAPs have limited regional coverage, which requires domestic coordination among MNOs. Satellites offer global coverage but demand cross-border or international cooperation, which is hindered by regulatory, privacy, and security concerns. Therefore, considering the above challenges, future research can focus on developing secure and efficient mechanisms for both domestic and international infrastructure sharing, as well as exploring the integration of privacy-preserving technologies such as blockchain.

\subsection{HIERARCHICAL ML-BASED OPTIMIZATION} 
\label{hierachical_ML_RICs}
\subsubsection{DEFINITION}
Hierarchical ML-based optimization refers to a multi-level learning and decision-making framework that decomposes complicated network control tasks across different timescales or functional layers. Within the ORAN architecture, intelligent control is inherently structured around the Non-RT RIC and Near-RT RIC, which naturally aligns with the hierarchical optimization framework across different timescales \cite{oranwhitepaper-1}. Prior studies have demonstrated the effectiveness of hierarchical ML for optimizing terrestrial ORAN systems in areas such as traffic steering \cite{Habib2025}, network slicing \cite{Ghafouri2024}, and resource allocation \cite{You2023}.

\subsubsection{RECENT ADVANCES AND PROMISING DIRECTIONS}
Several studies have explored the hierarchical ML-based optimization framework in NTN for different objectives \cite{shinde2024hierarchical, farajzadeh2025federated,umer2025intelligent}. The authors in \cite{farajzadeh2025federated} proposed a hierarchical federated learning (HFL) framework for NTNs, where HAPs constellations act as intermediate servers to enable global-scale training. \cite{shinde2024hierarchical} proposed a hierarchical reinforcement learning approach for joint network selection and computation offloading to achieve efficient resource utilization. \cite{umer2025intelligent} developed a hierarchical deep reinforcement learning (HDRL) framework for intelligent spectrum sharing in integrated TN-NTN networks, enhancing both spectral efficiency and network capacity. However, these proposed frameworks remain decoupled from the ORAN architecture.

In NTN scenarios, propagation delays can vary significantly across different platforms due to differences in altitude and orbital dynamics. In addition, processing capabilities are constrained by onboard payload hardware and the limited feeder link.
Traditional hierarchical optimization mechanisms, which typically assume uniform or bounded latency, are therefore insufficient. Therefore, an important future direction is to develop ORAN-based hierarchical ML solutions that can dynamically distribute intelligence across edge and cloud layers, and ensure efficient learning under the stringent latency constraints of NTNs.

\subsection{EDGE AI}\label{subsection_edge_AI}
\subsubsection{DEFINITION}
Edge AI typically refers to the integration of AI and ML capabilities at the network edge, enabling real-time decision-making and reducing the need for extensive data transmission \cite{letaief2021edge}. In future 6G networks, the integration of NTN and TN will rely on edge AI to support the emerging use cases that have stringent reliability requirements, such as autonomous driving, vehicular edge computing in V2X network, and virtual reality (VR) \cite{AI-for-TN-NTN-1}.

\subsubsection{RECENT ADVANCES AND PROMISING DIRECTIONS}
Recent studies have further explored the integration of edge AI into ORAN and NTN systems to enable the onboard decision making \cite{firouzi20252} and enhance the physical layer security \cite{chou2024edgeai}. The importance and the deployment design of edge AI in the NTN scenario are discussed in \cite{garcia2026edge}. In ORAN-based NTNs, one of the important future directions is to achieve intelligent partitioning of AI tasks between edge and cloud. This requires keeping latency-sensitive inference on NTN platforms, while offloading computation-heavy training to ground controllers. Meanwhile, since NTN platforms have stringent and different power constraints, another future direction is to focus on enhancing the energy efficiency of AI-driven RAN functions and control loops, which is essential for ensuring long-duration and cost-effective NTN operation.

In addition, LLMs have recently emerged as a powerful tool to improve the generalizability and adaptability of AI-driven network management \cite{salmi2025ai}. Recent studies have demonstrated the feasibility of integrating LLMs into the ORAN architecture or edge AI. The authors in \cite{tang2025end} introduced an edge AI orchestration framework that leverages LLM-based rApps for interactive and adaptive network service management. \cite{wu2025llm} proposed an ORAN system deploying LLM-empowered xApps in the Near-RT RIC to enable resilient, real-time decision-making. Traditional machine learning models, which rely primarily on offline training, often struggle to adapt to the highly dynamic and heterogeneous conditions of NTN environments. In contrast, LLM-based agents possess strong reasoning and contextual understanding capabilities, allowing them to respond adaptively to changing network conditions. Therefore, to achieve a more robust network management, future studies may focus on domain-specific fine-tuning of LLMs in rAPPs or xApps for NTN, as well as collaborative reasoning among distributed LLM agents in ORAN-based NTNs.

\subsection{CROSS-DOMAIN AI}
\subsubsection{DEFINITION}
Cross-domain AI, also known as multi-domain AI, is defined as the integration of collaborative AI-enabled functionalities across multiple domains \cite{oran-multidomain}. 
In wireless networks, this concept aims to jointly optimize network performance across RAN, TPN, and CN.
Cross-domain AI involves the integration of information, data exchange, collaborative model training and inference, as well as the sharing of computing resources. It also includes tasks such as power allocation in RAN, routing strategies in TPN, and slicing configurations in CN. By enabling coordinated decision-making across these domains, cross-domain AI can contribute to seamless multi-vendor integration and large-scale network optimization, thereby supporting scalable and resource-efficient ORAN infrastructures.

\subsubsection{RECENT ADVANCES AND PROMISING DIRECTIONS}
The proposed cross-domain SMO layer in Fig. \ref{Proposed_ORAN_NTN} inherently enables cross-domain AI capabilities. However, several challenges remain before fully exploiting the advantages of cross-domain AI. First, each domain has different optimization objectives. For instance, the RAN domain focuses on reliability and throughput, while the CN focuses on meeting SLAs for emerging applications such as AR/VR and real-time automation. Hence, designing an E2E optimization framework that balances these domain-specific objectives remains an important research direction. Second, the cross-domain AI controller must process heterogeneous data flows collected at different timescales. This necessitates the design of customized AI models that incorporate domain-specific knowledge in each data flow.

Moreover, in cross-domain AI, the open architectures and cross-domain data sharing may also introduce new vulnerabilities that compromise network security and privacy. To address these risks, several security mechanisms have been established. From a standardization perspective, the O-RAN Alliance’s Security Working Group (WG11) mandates zero trust architecture (ZTA) principles for all compliant components, enforcing mutually-authenticated transport layer security (mTLS), least-privilege access control, and standardized encryption, such as TLS 1.2/1.3 and SSH v2, to ensure secure data exchange across open interfaces \cite{oran-wg11}.
From the perspective of vendors or operators, they emphasize the secure software development lifecycle (SDLC) and the use of software bills of materials (SBOMs) when procuring and vetting third-party rApps and xApps \cite{nist-sbom}. To protect privacy, sensitive network data must be anonymized or aggregated by operators before being shared with external applications. In practice, the onboarding and validation process of these rApps and xApps would require containerization, sandboxing, and runtime monitoring to prevent any malicious application from compromising the network. These mechanisms collectively form a key security layer for operators running multivendor applications on the RIC.

Furthermore, AI models embedded in RICs, xApps, or rApps may also face adversarial attacks or model failures. To enhance the AI robustness or safety, future research can focus on advancing model verification techniques and resilient inference strategies. In addition, federated learning is being explored by major operators and vendors to optimize resource management in a privacy-preserving manner, without requiring the centralization of sensitive user data \cite{fl-privacy-2025}. However, its commercial-scale adoption for secure and large-scale network optimization is still at an early stage, highlighting the need for further research.

\subsection{MOBILITY MANAGEMENT}
\subsubsection{DEFINITION}
Mobility management is a critical challenge in NTNs, mainly encompassing handover, location management, and mobility robustness optimization. In TN, handover decisions are primarily coverage-driven and are typically triggered by the degradation of signal strength indicators such as RSRP and RSRQ. As the user equipment (UE) approaches the cell edge, the serving BS relies on UE measurement reports and predefined handover strategies to determine whether a handover should be initiated \cite{haghrah2023survey}.

However, in the satellite-based NTN, mobility is primarily driven by satellite beam movement rather than the UE's motion. For LEO satellites orbiting at 7.8 km/s, beams sweep rapidly across the Earth’s surface, causing even stationary users to experience frequent handovers every few minutes. This high handover frequency, together with large Doppler shifts, leads to increased UE power consumption and considerable signaling overhead \cite{alsaeedy2024survey}. In addition, the relatively constant UE-satellite distance within a beam footprint leads to minimal signal variations, making conventional measurement-based handover criteria less effective. Furthermore, inherent NTN characteristics such as long propagation delays, dynamic beam patterns, and intermittent feeder links may lead to outdated measurements and handover failures.

\subsubsection{RECENT ADVANCES AND PROMISING DIRECTIONS}
Recent studies have focused on improving mobility management in NTNs to ensure reliable connectivity \cite{alsaeedy2024survey,seeram2025handover,pasumarthy2026designing,3gpp38821}.
Considering the different payload deployment options in satellite-based NTN and the ORAN architecture, the handover procedures have been categorized into intra-DU (intra-satellite), inter-DU (inter-satellite), and inter-CU (inter-access) handovers in \cite{pasumarthy2026designing}. 3GPP has specified several NTN mobility management trigger mechanisms, based on measurement, location, time, timing advance, and elevation angle, respectively \cite{3gpp38821}. 
In \cite{alsaeedy2024survey}, the authors discussed different mobility management solutions for satellite-based 5G NTNs and evaluated their impacts on UE power consumption and signaling overhead, highlighting the potential of AI-based methods for dynamic mobility management optimization. The authors in \cite{seeram2025handover} focused on optimizing handover performance in NTN to enhance UE effective service time, and also investigated the impact of different ORAN functional splits between ground stations and LEO satellites on handover delays. This work revealed that the highest availability was achieved with the gNB onboard the satellite.

The above studies have advanced mobility management in NTNs from different perspectives. To achieve more effective mobility management within the proposed ORAN-based NTN framework, several research directions can be further explored. First, ephemeris-aware xApps should be developed to support proactive beam and satellite handovers, and Near-RT RIC policies need to be designed to mitigate Doppler and propagation delay effects \cite{baranda2025architectural}. Second, AI/ML models can be leveraged to explore the multi-constellation mobility mechanisms for seamless connectivity. In addition, lightweight inference frameworks are also required to ensure reliable decision-making under the stringent backhaul constraints.

\subsection{DIGITAL TWIN}
\subsubsection{DEFINITION}
The digital twin (DT) is a virtual replica of a physical system that enables simulation, analysis, and optimization of system behavior in a risk-free environment \cite{tran2025network}. 
For telecom companies, the direct testing of new network setups in operational networks usually involves high risks, since its failures may lead to service outages, significant revenue loss, and reputational damage. As a result, new network configurations are often evaluated in controlled laboratory environments with limited data and a small number of users. This approach, commonly referred to as a pilot environment, enables service providers to assess the interaction of new technologies with existing systems, without risking large-scale customer disruptions. To provide realistic system evaluation while further mitigating operational risks, implementing DT in wireless networks has become essential for operators and users. 

In the proposed ORAN-based NTN framework, the integration of DTs helps operators implement different regenerative architectures and evaluate E2E performance under diverse wireless channels. In addition, DTs also further empower ORAN to perform network analytics and diagnosis for high-efficiency service in highly dynamic NTN scenarios \cite{zhou2025digital}.

\subsubsection{RECENT ADVANCES AND PROMISING DIRECTIONS}

DTs are envisioned as a key enabler for intelligent network design, analysis, operation, and automation. They complement the AI-Native management frameworks in future 6G networks. Recent research has highlighted the growing importance of AI, DT, and ORAN for 6G NTN scenarios \cite{campoy2025digital,wu2024ai,tao2024wireless,moore2025next,mushi2024open,Hraishawi2025,firouzi20252}. Specifically, several studies have explored AI-driven approaches to enhance DT accuracy and scalability. For instance, reinforcement learning (RL) has been integrated into scalable 6G RAN DTs to enable adaptive decision-making in \cite{campoy2025digital}, and the authors in \cite{wu2024ai} highlighted the integration of ML, deep learning (DL), and natural language processing (NLP) in DTs. Generative AI (GenAI) models such as transformers and diffusion architectures have been employed to construct hierarchical DT frameworks in \cite{tao2024wireless}. In addition, DT-based emulation platforms and testbeds have been developed to support ORAN-based NTN experimentation in \cite{moore2025next,mushi2024open}. The DT-enabled ORAN framework has been investigated in terms of its effectiveness in optimizing resource allocation \cite{Hraishawi2025} and facilitating mission-aware autonomy in dynamic NTN environments \cite{firouzi20252}.

As discussed above, ORAN enables on-orbit tests and upgrades for NTN and thus avoids the time-consuming launch process. However, on-orbit tests and upgrades in the real and complicated NTN environment face much higher risks than traditional TN, which highlights the need for high-fidelity DT technologies \cite{liu2024challenges}. 
Within the proposed framework, the Near-RT RIC has the AI inference capability to create a virtual replica of an ORAN-based NTN, but also faces several challenges. Specifically, to guarantee the inference accuracy, the Near-RT RIC needs to maintain strict synchronization with the physical network elements O-RU/O-DU/O-CU under this highly mobile scenario. In addition, unlike the traditional terrestrial DT with adequate communication and computation capabilities in the central server, the Near-RT RIC in NTNs must also guarantee the xApps execution and the existing communication overhead with other network components. To address the challenges above, a hierarchical DT framework should be developed for ORAN-based NTN. 
To be more specific, a DT of Near-RT RIC is first established to model and predict its mobility as well as the available communication and computation resources. Based on these predictions, the Near-RT RIC subsequently schedules the available resources to construct a system-level DT of the ORAN-based NTN for inference tasks.

\section{CONCLUSION}
\label{conclusion}
Developing efficient and flexible NTN architectures is essential for AI-based wireless networks to achieve ubiquitous coverage and support advanced 6G use cases. This survey has presented a comprehensive and structured overview related to the development of the AI-Native ORAN-based NTN framework, supporting dynamic configuration, scalability, and intelligent orchestration in future 6G NTN systems.
Specifically, this survey first reviews the state-of-the-art research on ORAN-based NTNs, highlighting their design principles, technical progress, and existing limitations. It also offers necessary background knowledge and key related literature on ORAN, NTN, and AI-Native for communications. Furthermore, this paper analyzes the unique challenges encountered across the DevOps lifecycle of NTNs and proposes an orchestrated AI-Native ORAN-based NTN framework. Meanwhile, the architecture and key technological enablers of the proposed framework, including dynamic wireless FH splits, multi-tier RIC structure, flexible functional role switch, and cross-domain SMO, are discussed in detail. More importantly, this paper presents several representative use cases to demonstrate the applicability of the proposed framework. Finally, this paper outlines prospective future research directions to further enhance the AI-Native network control, operation, and optimization in NTN.




\section*{ACKNOWLEDGMENT}
The authors would like to thank the support of the interdisciplinary research center for communication systems and sensing (IRC-CSS), King Fahd University of Petroleum and Minerals (KFUPM). The authors would also like to thank the support of the Office of Sponsored Research, King Abdullah University of Science and Technology (KAUST).

\ifCLASSOPTIONcaptionsoff
  \newpage
\fi

\bibliographystyle{ieeetr} 
\bibliography{Bibliography}

\end{document}